\documentclass[12pt,unsortedaddress,eqsecnum,amsmath,amssymb,aip,jcp,preprint,floatfix]{revtex4-1}
\usepackage{graphicx}
\usepackage{color}
\usepackage[version=3]{mhchem}
\usepackage{dcolumn}
\usepackage{multirow}
\usepackage[normalem]{ulem}
\usepackage{cmbright}
\usepackage[dvipsnames]{xcolor}

\usepackage{comment}

\begin{document}

\newcommand\bfec[1]{\textcolor{blue}{\textit{ XXX BFEC: #1 XXX }}}
\newcommand\jtt[1]{\textcolor{orange}{\textit{ XXX JTT: #1 XXX }}}
\newcommand\djt[1]{\textcolor{green}{\textit{ XXX DJT: #1 XXX }}}

\newcommand{\g}[2]{\mathbf{g}_{#1#2}(\R)}
\newcommand{\h}[2]{\mathbf{h}_{#1#2}(\R)}
\newcommand{\nacv}[2]{\mathbf{d}_{#1#2}(\R)}
\newcommand{\R}{\mathbf{R}}

\newcommand{\xua}[2]{\bar{\mathbf{x}}_{#1#2}(\R)}
\newcommand{\yua}[2]{\bar{\mathbf{y}}_{#1#2}(\R)}
\newcommand{\xa}[2]{\bar{\mathbf{x}}'_{#1#2}(\R)}
\newcommand{\ya}[2]{\bar{\mathbf{y}}'_{#1#2}(\R)}

\newcommand{\xual}[2]{\bar{x}_{#1#2}}
\newcommand{\yual}[2]{\bar{y}_{#1#2}}
\newcommand{\xal}[2]{\bar{x}'_{#1#2}}
\newcommand{\yal}[2]{\bar{y}'_{#1#2}}

\setlength{\tabcolsep}{12pt}

\DeclareFontFamily{OT1}{cmbr}{\hyphenchar\font45 }
\DeclareFontShape{OT1}{cmbr}{m}{n}{%
  <-9>cmbr8
  <9-10>cmbr9
  <10-17>cmbr10
  <17->cmbr17
}{}
\DeclareFontShape{OT1}{cmbr}{m}{sl}{%
  <-9>cmbrsl8
  <9-10>cmbrsl9
  <10-17>cmbrsl10
  <17->cmbrsl17
}{}
\DeclareFontShape{OT1}{cmbr}{m}{it}{%
  <->ssub*cmbr/m/sl
}{}
\DeclareFontShape{OT1}{cmbr}{b}{n}{%
  <->ssub*cmbr/bx/n
}{}
\DeclareFontShape{OT1}{cmbr}{bx}{n}{%
  <->cmbrbx10
}{}

\title{On the description of conical intersections between excited electronic states with LR-TDDFT and ADC(2)}
\author{Jack T. Taylor}
\author{David J. Tozer}
\email{d.j.tozer@durham.ac.uk}
\affiliation{Department of Chemistry, Durham University, South Road, Durham DH1 3LE, UK}
\author{Basile F. E. Curchod}
\email{basile.curchod@bristol.ac.uk}
\affiliation{Centre for Computational Chemistry, School of Chemistry, University of Bristol, Cantock's Close, Bristol BS8 1TS, UK}
\date{\today}
\begin{abstract}
Conical intersections constitute the conceptual bedrock of our working understanding of ultrafast, nonadiabatic processes within photochemistry (and photophysics). Accurate calculation of potential energy surfaces within the vicinity of conical intersections, however, still poses a serious challenge to many popular electronic structure methods. Multiple works have reported on the deficiency of methods like linear-response time-dependent density functional theory within the adiabatic approximation (AA LR-TDDFT) or algebraic diagrammatic construction to second-order (ADC(2)) -- approaches often used in excited-state molecular dynamics simulations -- to describe conical intersections between the ground and excited electronic states. In the present study, we focus our attention on conical intersections \textit{between excited electronic states} and probe the ability of AA LR-TDDFT and ADC(2) to describe their topology and topography, using protonated formaldimine and pyrazine as two exemplar molecules. We also take the opportunity to revisit the performance of these methods in describing conical intersections involving the ground electronic state in protonated formaldimine -- highlighting in particular how the \textit{intersection ring} exhibited by AA LR-TDDFT can be perceived either as a (near-to-linear) seam of intersection or two interpenetrating cones, depending on the magnitude of molecular distortions within the branching space.
\end{abstract}

\maketitle

\section{Introduction}
\label{intro}

A theoretical understanding of almost all chemical processes arguably stems from the fundamental concept of static potential energy surfaces (PESs), a consequence of invoking the Born-Huang representation\cite{born_dynamical_1988} for the molecular wavefunction. Of particular significance to photochemical (and photophysical) processes is the notion of conical intersections (CXs), which correspond to molecular geometries where two (or more) adiabatic PESs become energetically degenerate.\cite{hettema_behaviour_2000, teller_crossing_1937, herzberg_intersection_1963} In contrast to initial opinions,\cite{ van_der_lugt_symmetry_1969, michl_physical_1974} it is now agreed\cite{truhlar_relative_2003} that CXs are far from arcane mathematical curiosities. Instead, they play a critical mechanistic role in our theoretical framework to understand the ultrafast, nonradiative decay from the excited electronic states of a molecule to its ground electronic state.\cite{levine_isomerization_2007, yarkony_nonadiabatic_2012, matsika_nonadiabatic_2011, boeije_one-mode_2023} Uncovering the pivotal influence of CXs within photochemistry has triggered a plethora of works, both from an applied and theoretical perspective.\cite{riad_manaa_noncrossing_1990, bernardi_mechanism_1990, xantheas_potential_1991, atchity_potential_1991, polli_conical_2010, martinez_seaming_2010}

Formally, CXs only appear when using an adiabatic electronic basis (i.e., the eigenstates of the electronic Hamiltonian) within the Born-Huang representation\cite{born_dynamical_1988} of the molecular wavefunction.\footnote{In other words, if either a diabatic electronic basis or the exact-factorisation representation\cite{abedi_exact_2010, abedi_correlated_2012, gonzalez_exact_2020} of the molecular wavefunction were used instead, CXs would not be observed.\cite{curchod_dynamics_2017, agostini_when_2018, ibele_photochemical_2022}} CXs do not exist for isolated molecular geometries, but instead (for a CX between two states) they comprise an ($F - 2$)-dimensional seam (or intersection) space (where $F = 3N - 6$ nuclear degrees of freedom for a non-linear molecule with $N$ atoms) and an orthogonal two-dimensional branching\cite{atchity_potential_1991} (or $g-h$)\cite{yarkony_conical_2001} space. In particular, the branching space is spanned by two vectors that depend on the nuclear coordinates, $\R$: the gradient difference vector, $\g{i}{j}$, and the derivative coupling vector, $\h{i}{j}$, where $i$ and $j$ denote electronic states. Movement along these two vectors lifts the energy degeneracy, doing so linearly, giving the characteristic double-cone topology within the branching space.\cite{yarkony_diabolical_1996} However, distortion along any of the remaining $F - 2$ nuclear degrees of freedom retains the degeneracy and the geometry remains within the seam space. Moreover, the local minima within the seam space -- termed minimum-energy CXs (MECXs) -- are typically used to characterise the nonadiabatic transitions between the electronic states.\cite{levine_optimizing_2008}

As always, the insolubility of the exact electronic Schrödinger equation for chemically relevant systems necessitates using approximate electronic structure methods. Whether a given electronic structure method can adequately predict the \textit{topology} (i.e., the dimensionality of the CX branching or seam spaces) and the \textit{topography} (i.e., the shape of the PESs in the vicinity of the CX point within the branching space) of a given CX is a key consideration in nonadiabatic molecular dynamics simulations.\cite{gozem_shape_2014} Much attention has therefore been paid to benchmarking different electronic structure methods in this context -- see Ref. \onlinecite{matsika_electronic_2021} for a recent review. Two requirements are often highlighted as being critical for an accurate description of CXs involving the ground electronic state: (i) inclusion of dynamic and static electron correlation, given that the character of the electronic states changes rapidly in the vicinity of (and passing through) a CX; (ii) a balanced treatment of the ground and excited electronic states, so as to allow explicit coupling between them.\cite{gozem_shape_2014, liu_analytical_2021} The obvious electronic structure methods of choice have thus been multiconfigurational and multireference methods\cite{granovsky_extended_2011, gozem_dynamic_2012, gozem_shape_2014, sen_comprehensive_2018, park_analytical_2020, battaglia_role_2021, nishimoto_analytic_2022} such as MCSCF and MRCI, with the state-averaged complete active space self-consistent field (SA-CASSCF)\cite{roos_complete_1980} approach being the most widely used. A popular alternative that extends upon SA-CASSCF by including a more balanced description of dynamic correlation, which has seen a recent rise in use within excited-state molecular dynamics simulations,\cite{park_--fly_2017, sen_comprehensive_2018, polyak_ultrafast_2019, heindl_xms-caspt2_2019, park_single-state_2019, winslow_comparison_2020, chakraborty_effect_2021, zhang_nonadiabatic_2022} is extended multi-state complete active space second-order perturbation theory (XMS-CASPT2).\cite{andersson_second-order_1990, andersson_secondorder_1992, finley_multi-state_1998, granovsky_extended_2011, shiozaki_communication_2011}

Given the high computational cost of multiconfigurational and multireference methods and the ever-increasing size of the systems to which they need to be applied, cheaper alternatives to add to the photochemists' toolkit are still in demand. Using simpler, single-determinant methods -- often designed for calculations of excited electronic states within the Franck-Condon (FC) region -- to describe CXs between the \textit{ground} and \textit{excited electronic states} has, however, proven problematic. Notable examples include linear-response time-dependent density functional theory within the adiabatic approximation (AA LR-TDDFT),\cite{runge_density-functional_1984, chong_time-dependent_1995, petersilka_excitation_1996, ullrich_time-dependent_2011} algebraic diagrammatic construction (ADC) methods\cite{schirmer_beyond_1982, trofimov_efficient_1995, dreuw_algebraic_2015, hattig_structure_2005} and coupled cluster theories.\cite{stanton_equation_1993, comeau_equation--motion_1993, christiansen_second-order_1995, krylov_size-consistent_2001, krylov_spin-flip_2006, krylov_equation--motion_2008, sneskov_excited_2012, tuna_assessment_2015} In particular, AA LR-TDDFT (within the Tamm-Dancoff approximation (TDA)\cite{hirata_time-dependent_1999}) and ADC(2) have been thoroughly tested due to the appeal of using these low-cost electronic structure methods within nonadiabatic molecular dynamic simulations.\cite{tapavicza_trajectory_2007, werner_nonadiabatic_2008, curchod_trajectory-based_2013, plasser_surface_2014, crespo-otero_recent_2018} 

In contrast, little is known\cite{levine_conical_2006} about the precise quality of these cheaper approaches in describing CXs \textit{between excited electronic states}. Although considering electronic energies alone may suggest an adequate representation of CXs within AA LR-TDDFT/TDA and ADC(2) in this context, is this what one observes in practice? How well do the topology and topography of CXs between excited electronic states given by these single-determinant methods reproduce those predicted by multiconfigurational and multireference techniques?

The present study attempts to address these questions from a pragmatic perspective by investigating the ability of AA LR-TDDFT/TDA and ADC(2) to describe CXs between the lowest two excited singlet electronic states, S$_1$ and S$_2$, for two exemplar molecules, protonated formaldimine and pyrazine. We also revisit the problem faced by AA LR-TDDFT/TDA in describing CXs between the ground electronic state, S$_0$, and S$_1$ for the case of protonated formaldimine, focusing on the behaviour of the PESs within the branching space at varied distances away from the MECX geometry. Despite providing a static, electronic structure perspective in this work, we bear nonadiabatic molecular dynamics in mind, choosing to compare our AA LR-TDDFT/TDA and ADC(2) results to reference XMS-CASPT2 results. Our work is organised as follows: We start by (i) reviewing the problem of CXs involving the ground electronic state from AA LR-TDDFT and considering issues relevant to CXs between excited states, before (ii) presenting the computational details of our calculations. We then (iii) explore the S$_2$/S$_1$ and S$_1$/S$_0$ MECX branching spaces of protonated formaldimine as predicted by the three electronic structure methods, followed by (iv) the S$_2$/S$_1$ MECX of pyrazine, where further considerations of the exchange-correlation functional used in AA LR-TDDFT/TDA are provided.

\section{Methods}
\subsection{Notes on the description of conical intersections with AA LR-TDDFT}
\label{tddft_vs_cxs}

The inaccurate description of PESs in the vicinity of CXs involving the ground electronic state is, by now, a well-reported deficiency of LR-TDDFT within the AA. The first investigation to highlight this problem was that of Levine \textit{et al.},\cite{levine_conical_2006} where for linear H$_2$O the dimensionality of the intersection was shown to be $F-1$ rather than $F-2$ (i.e., incorrect topology), whilst for H$_3$ the shape of the first excited-state PES was shown to vary too rapidly near the intersection point (i.e., incorrect topography), despite the CX possessing the correct dimensionality. Tapavicza \textit{et al.}\cite{tapavicza_mixed_2008} subsequently showed that applying the TDA not only helps to reduce excited-state instability problems, but also gives an approximate S$_1$/S$_0$ CX for oxirane with a slightly interpenetrating double cone. Further studies have provided additional examples of the issues of AA LR-TDDFT in describing CXs between the ground and first excited electronic states, e.g., see Refs.~\citenum{gozem_shape_2014, huix-rotllant_assessment_2013, ferre_description_2015, ferre_surface_2014, marques_non-bornoppenheimer_2012}. We note, however, that AA LR-TDDFT has been shown to predict reasonably accurate S$_1$/S$_0$ CX geometries and branching planes, despite issues with the PESs.\cite{levine_conical_2006}

A common starting point for analysing the deficiencies of AA LR-TDDFT is to consider the description of CXs involving the ground state within the alternative (wavefunction) approach of configuration interaction singles (CIS). Like AA LR-TDDFT, CIS (i) uses a single Slater determinant as its reference and (ii) comprises a set of linear equations restricted to a single-excitation subspace. In CIS, the coupling (i.e., Hamiltonian matrix elements) between the ground and excited electronic states is zero for any molecular geometry by virtue of Brillouin’s theorem;\cite{szabo_modern_1996} this renders one of the two conditions for electronic degeneracy\cite{zhang_nonadiabatic_2021, hettema_behaviour_2000} at a CX to be satisfied trivially for any nuclear configuration — i.e., the corresponding derivative coupling vector, $\h{0}{1}$, in CIS is zero. As a result, CIS exhibits a linear ($F-1$)-dimensional intersection (as opposed to a conical ($F-2$)-dimensional intersection), where the degeneracy is only lifted along one (not both) branching space vector direction(s).\cite{levine_conical_2006, tapavicza_mixed_2008} Given the CIS excited state and Hartree-Fock (HF) reference state do not ‘see each other’ due to the lack of coupling,\cite{tapavicza_mixed_2008} their corresponding PESs cross each other within the branching space, leading to regions where the CIS excited state becomes lower in energy than the HF reference state (i.e., one observes negative excitation energies). The HF reference state struggles to reproduce the necessary rapid change in electronic character near the CX.\cite{levine_conical_2006}

Despite the similarity between the approaches, these CIS arguments cannot be used to explain why AA LR-TDDFT fails to correctly describe CXs between the ground and excited electronic states. This is because Brillouin's theorem does \textit{not} hold within (LR-TD)DFT\cite{tavernelli_non-adiabatic_2009, tapavicza_mixed_2008, ullrich_time-dependent_2011} because the method does not provide formal access to wavefunctions (only electron densities). The Kohn-Sham (KS) determinant is the wavefunction of the non-interacting system, not the interacting system. Similarly, while excited-state wavefunctions can be reconstructed using excited Kohn-Sham determinants (for electronic state assignment purposes -- see Ref.~\onlinecite{chong_time-dependent_1995}), they do not correspond to excited-state wavefunctions of the interacting system. The situation is reminiscent of the calculation of $\langle S^2 \rangle$ / spin contamination in DFT, whereby the usual single determinant expression is not appropriate for the interacting system.\cite{cohen_evaluation_2007, pople_spin-unrestricted_1995} In spite of the absence of Brillouin's theorem, it is still argued\cite{matsika_electronic_2021, yang_conical_2016} that there is no coupling between the ground and excited states in AA LR-TDDFT and so the method is expected to exhibit \textit{similar} CX problems to CIS. This lack of coupling in LR-TDDFT is a consequence of using the adiabatic approximation (as well as the ground-state exchange-correlation functional approximation). Within AA LR-TDDFT, the ground (reference) state is variationally obtained within an initial DFT calculation, separate to the singly-excited (response) states, which are obtained when the Casida equation is solved (i.e., $E_j(\boldsymbol{R}) = E_0(\boldsymbol{R}) + \omega_j(\boldsymbol{R})$, where $\omega_j(\boldsymbol{R}$) is the $j^{th}$ vertical excitation energy).\cite{herbert_spin-flip_2022} The ground and excited states are therefore not treated on an equal footing, and so the coupling between them is absent. We note, this is the same reason why ADC(2) struggles to accurately predict CXs involving the ground state -- the ground state is obtained at the MP2 level of theory, whereas the excited states are obtained with ADC(2).\cite{tuna_assessment_2015} 

Many attempts have been made to fix (or, at least, circumvent) the incorrect description of CXs involving the ground electronic state within AA LR-TDDFT; these approaches can be broadly divided into two categories: (i) those that artificially expand the dimension of the LR-TDDFT(/TDA) problem to introduce coupling between the ground and excited states and (ii) those rooted solely within the formal linear response framework of TDDFT. For the first category, methods either incorporate explicit double excitations\cite{teh_simplest_2019, athavale_inclusion_2021, ferre_many-body_2015} (since these introduce coupling between the ground and excited states within a configuration interaction picture, improving upon CIS), or include direct coupling between the reference KS determinant and (at least one) singly-excited determinant(s).\cite{li_configuration_2014, shu_dual-functional_2017, shu_dual-functional_2017-1, kaduk_communication_2010} Some fulfill this goal by using DFT quantities in a larger CI-type matrix, interpreting Slater determinants constructed from KS orbitals as approximations to the real, interacting wavefunctions,\cite{li_configuration_2014, teh_simplest_2019, athavale_inclusion_2021} whilst others add selected excited contributions to the AA LR-TDDFT/TDA matrix equations from those derived within many-body perturbation theory.\cite{ferre_many-body_2015} The second category of methods instead comprise different variants of standard LR-TDDFT; they generate, via a modified linear response formalism, the ground and excited states of interest \textit{together} as response states from a sacrificial reference state\cite{zhang_nonadiabatic_2021, herbert_density-functional_2023, herbert_spin-flip_2022} while still preserving the AA. These methods include spin-flip TDDFT,\cite{bernard_general_2012, krylov_size-consistent_2001, shao_spinflip_2003, herbert_beyond_2016} particle-particle RPA(/TDA)\cite{van_aggelen_exchange-correlation_2013, yang_double_2013, yang_excitation_2014, yang_conical_2016} and hole-hole TDA\cite{bannwarth_holehole_2020, yu_ab_2020} and, in all cases, the resulting ground and excited states are treated on the same footing.

The aforementioned approaches are pragmatic. However, the ultimate goal within conventional LR-TDDFT is to rigorously go beyond the AA by using a frequency-dependent exchange-correlation kernel. In the exact case, the LR-TDDFT matrix problem represents a set of non-linear equations\footnote{We note, this is in contrast to the situation in CIS, where the equations are always linear and only ever approximate (i.e., the exact case would be full CI). Even within the AA, where now the LR-TDDFT equations are indeed linear, the response matrix elements in AA LR-TDDFT(/TDA) differ from that in CIS (Ref.~\citenum{dreuw_single-reference_2005}), where the former depends on the response of the multiplicative exchange-correlation potential, whereas the latter depends on the response of the non-multiplicative HF exchange potential.} that, despite being built in a basis of single excitations, have folded in all the information from double and higher (de-)excitations thanks to the frequency dependence of the exact exchange-correlation kernel.\cite{ferre_many-body_2015, authier_dynamical_2020, romaniello_double_2009} It could be argued (i.e., along similar lines to comments made by Huix-Rottlant and Casida in Ref. \onlinecite{ferre_many-body_2015}) that a combination of these single, double and higher (de-)excitations from the DFT reference state (i.e., a single KS determinant) could lead to the true correlated ground state being reproduced in the linear-response excitation manifold along with the (similarly correlated) excited states.\cite{maitra_undoing_2005, maitra_long-range_2006} The ground and excited electronic states would then, therefore, be treated on an equal footing, establishing the required coupling between them. 

We now address a less frequently asked question: how well does AA LR-TDDFT perform for CXs between excited electronic states? Given that excited states are treated on an equal footing within LR-TDDFT (i.e., they are obtained \textit{together} when one solves the Casida equation), it may be expected that, even in the AA, the coupling between respective excited states is indeed present. As a result, the aptitude of AA LR-TDDFT to correctly predict the topology and topography of CXs between excited electronic states is often taken for granted, even if little (in the way of explicit plotting of the excited-state CX branching spaces) is known about the performance of the method in this context. \footnote{It should be noted that Levine et al. did provide a very brief discussion about the description of CXs between excited states with AA LR-TDDFT for molecules, such as malonaldehyde and benzene, within their seminal work (Ref.~\citenum{levine_conical_2006}). However, no explicit branching space plots were presented.} We note that the same also applies to excited electronic states obtained with ADC(2). One aspect, in particular, that requires attention when discussing CXs between excited electronic states with LR-TDDFT is the description of the branching space vectors, especially the derivative coupling vector, $\h{i}{j}$ (and, by extension, the closely related (first-order) nonadiabatic coupling vector, $\nacv{i}{j}$, where $\h{i}{j} = [E_j(\R) - E_i(\R)] \times \nacv{i}{j}$). The $\h{i}{j}$ vectors between the ground and excited electronic states are well-defined in linear-response TDDFT and can be derived from the excited electronic density.\cite{chernyak_density-matrix_2000, baer_introduction_2002, tapavicza_trajectory_2007, send_first-order_2010, hu_nonadiabatic_2007, tavernelli_nonadiabatic_2009-1, tavernelli_nonadiabatic_2009} These $\h{0}{1}$ vectors are formally exact within the limit that LR-TDDFT, itself, becomes exact (i.e., beyond the AA and when using the exact ground-state exchange-correlation functional), and only become approximate when the aforementioned approximations are invoked. This contrasts with the $\h{0}{1}$ vectors in CIS, which, as already mentioned, are formally zero by definition. On the other hand, the $\h{i}{j}$ vectors between excited electronic states can be defined in CIS, but their quality depends on the accuracy of the underlying CIS level of theory used to describe the coupled electronic states. The situation is different for LR-TDDFT, as even in the exact case, the $\nacv{i}{j}$ vectors (and therefore the $\h{i}{j}$ vectors) between excited electronic states can formally only ever be approximate within a linear-response formalism -- quadratic-response is required to derive an exact expression.\cite{tavernelli_nonadiabatic_2010, li_first-order_2014, li_first_2014, wang_nac-tddft_2021, parker_multistate_2019} While numerical test indicate that $\h{i}{j}$ vector between excited electronic states might be fairly well-approximated within a linear-response formalism,\cite{tavernelli_nonadiabatic_2010,ou_derivative_2015} in particular within the TDA, a proper description of the branching space for CXs between excited electronic states is far from granted within AA LR-TDDFT, despite its routine use in excited-state dynamics simulations involving multiple excited electronic states. This work hopes to provide some reassurance on the behaviour of AA LR-TDDFT/TDA (and ADC(2)) for CXs involving two excited electronic states.

\subsection{Computational details}
\subsubsection{Electronic structure}
\label{comp_dets_ele_struct}

All XMS-CASPT2 energies, energy gradients\cite{vlaisavljevich_nuclear_2016} and nonadiabatic coupling vectors\cite{park_analytical_2017} were determined with the BAGEL 1.2.0 program package.\cite{shiozaki_span_2018} The single-state, single-reference (SS-SR) contraction scheme\cite{vlaisavljevich_nuclear_2016, gonzalez_multiconfigurational_2020} was employed for all XMS-CASTP2 calculations with a real vertical shift of 0.3 a.u. to avoid intruder state issues. Density fitting and frozen core approximations were also applied. For protonated formaldimine, a three-state averaging and a (6/4) active space, comprising the two pairs of C-N $\sigma\sigma^*$ and $\pi\pi^*$ orbitals (Fig. S1a), were used (following Ref.~\citenum{barbatti_ultrafast_2006}). For pyrazine, a three-state averaging and a (10/8) active space, including the six $\pi$ orbitals and two nitrogen lone pairs (Fig. S1b), were employed (based on Ref.~\citenum{shiozaki_pyrazine_2013}). All DFT\cite{hohenberg_inhomogeneous_1964, kohn_self-consistent_1965, parr_density-functional_1989} and AA LR-TDDFT/TDA energies, energy gradients and nonadiabatic coupling vectors were determined with a development version of the GPU-accelerated TeraChem 1.9 program package.\cite{isborn_excited-state_2011, ufimtsev_quantum_2008, ufimtsev_quantum_2009, ufimtsev_quantum_2009-1, titov_generating_2013, seritan_terachem_2020, seritan_span_2021} The PBE0 (global hybrid) exchange-correlation functional\cite{perdew_generalized_1996, adamo_toward_1999, ernzerhof_assessment_1999} was used throughout (unless otherwise stated -- see the SI) within the TDA. All MP2\cite{moller_note_1934} and ADC(2) energies and energy gradients\cite{hattig_distributed_2006, hattig_structure_2005} were determined with the Turbomole 7.4.1 program package,\cite{ahlrichs_electronic_1989, furche_turbomole_2014} employing frozen core and resolution of identity\cite{weigend_efficient_2002} approximations. The Dunning cc-pVTZ basis set was used in all XMS-CASPT2, MP2 and ADC(2) calculations, whereas the Dunning cc-pVDZ basis set was used in all DFT and AA LR-TDDFT/TDA calculations.\cite{dunning_gaussian_1989} The density fitting procedure, utilised in all XMS-CASPT2 calculations, made use of the cc-pVTZ-jkfit auxiliary basis set from the BAGEL library. 
For clarity, we will drop the 'AA' hereafter when discussing our LR-TDDFT/TDA results. For quantities involving excited states only, we use the notation LR-TDDFT/TDA/PBE0 and ADC(2). For quantities involving ground and excited states, we use the notation (LR-TD)DFT/TDA/PBE0 and MP2/ADC(2).

\subsubsection{Critical geometries and linear interpolation in internal coordinates}
\label{comp_dets_geoms+liic}

\textbf{Protonated formaldimine}. The S$_0$ minimum (commonly denoted FC), S$_2$/S$_1$ MECX and S$_1$/S$_0$ MECX geometries were first optimised with XMS-CASPT2. MECX geometry optimisation utilised the gradient-projection algorithm of Bearpark \textit{et al}.\cite{bearpark_direct_1994} Linear interpolation in internal coordinates (LIIC) pathways were generated to connect these three critical geometries of protonated formaldimine. An LIIC pathway serves as the most direct way of connecting two key points in configurational space by interpolating new points based on internal (rather than Cartesian) coordinates;\cite{hudock_ab_2007} as such, they do not constitute minimum-energy pathways. A single-point XMS-CASPT2 energy calculation was performed for each geometry to obtain the three lowest electronic states, S$_0$, S$_1$, and S$_2$, along the LIIC. Electronic energies are given relative to the S$_0$ energy at the S$_0$ minimum. 

The same procedure was repeated to acquire the electronic energies along corresponding LIIC pathways for (LR-TD)DFT/TDA/PBE0 and for MP2/ADC(2), respectively. As noted in Section \ref{tddft_vs_cxs}, neither (LR-TD)DFT/TDA, nor MP2/ADC(2) adequately describe the branching space of S$_1$/S$_0$ CXs. Therefore, we use the term minimum-energy \textit{crossing points} (MECPs) instead of minimum-energy \textit{conical intersections} (MECXs) when referring to the S$_1$/S$_0$ intersection geometries located upon applying MECX optimisation algorithms with these two electronic structure methods. To locate the MECXs (or MECPs) with (LR-TD)DFT/TDA or MP2/ADC(2), we used a combination of different geometry optimisation algorithms to ensure that the lowest possible electronic energy was found for these critical points. For (LR-TD)DFT/TDA, the gradient-projection method of Bearpark \textit{et al.},\cite{bearpark_direct_1994} the Lagrange-Newton method of Manaa and Yarkony,\cite{manaa_intersection_1993} the penalty-function of Ciminelli \textit{et al.}\cite{ciminelli_photoisomerization_2004} and the CIOpt method of Levine \textit{et al.}\cite{levine_optimizing_2008} were used; CIOpt was used for MP2/ADC(2) with subsequent refinement of the MECX (or MECP) geometries carried out within their respective branching spaces. The details of these procedures can be found in the SI.

It is important to stress here that in each case, the same electronic structure method was used to calculate the electronic energies and to optimise the three critical geometries.

\textbf{Pyrazine.} The same procedure was used to optimise the critical geometries in pyrazine. Only the S$_0$ minimum and S$_2$/S$_1$ MECX geometries were considered using the three electronic structure methods. Equally, we do not present LIIC plots for pyrazine.

\subsubsection{Plotting the CX branching space}
\label{comp_dets_branch_space}

The branching space vectors, $\g{i}{j}$ and $\h{i}{j}$, were first computed using XMS-CASPT2 at the optimised XMS-CASPT2 S$_j$/S$_i$ MECX geometry. The branching space vectors were then orthogonalised by the Yarkony procedure\cite{yarkony_adiabatic_2000, yarkony_conical_2001} and appropriately normalised, before being used to generate a 2D grid of 29$\times$29 geometries along the branching plane, centred on the optimised XMS-CASPT2 S$_j$/S$_i$ MECX geometry. To facilitate this, nuclear distortions along the orthonormalised $\xua{i}{j}$ and $\yua{i}{j}$ vector directions (see SI for branching space vector definitions) were multiplied by an appropriate scale factor and added in fourteen increments in the positive and negative directions, respectively, spanning $\pm 0.001$ a.u. in both branching space vector directions, as was done similarly in Ref. \onlinecite{lefrancois_spin-flip_2017}. At each grid-point geometry, a single-point XMS-CASPT2 energy calculation was performed, giving the S$_i$ and S$_j$ PESs in the region surrounding the optimised XMS-CASPT2 S$_j$/S$_i$ MECX geometry. Electronic energies are given relative to the S$_i$ energy at the MECX geometry, which is located at the grid origin.

The same procedure was repeated to obtain the corresponding S$_j$/S$_i$ MECX (or MECP) branching spaces of (LR-TD)DFT/TDA/PBE0 and MP2/ADC(2), respectively. For direct comparison of the branching space plots in Figs. \ref{fig:pro_form_s2s1_bp}-\ref{fig:pyra_s2s1_bp} (and Figs. S2, S4, S5 and S7 in the SI) obtained by the different electronic structure methods, we followed the approach taken in Ref. \onlinecite{liu_analytical_2021}: the orthonormalised branching space vectors were rotated within their respective branching planes to ensure maximal overlap with the reference orthonormalised vectors of XMS-CASPT2. These new rotated (orthonormalised) branching space vectors are denoted $\xa{i}{j}$ and $\ya{i}{j}$. Details of this rotation procedure and the process used to orthonormalise the raw branching space vectors are provided in the SI.

We stress again that in each case, the same electronic structure method was used to compute the electronic energies, branching space vectors and to optimise the MECX (or MECP) geometries, except for MP2/ADC(2), where the $\h{i}{j}$ vector from XMS-CASPT2 was used instead. Therefore, the branching spaces constructed are fully-consistent (where possible) within each electronic structure method.

\section{Results and discussion}
\subsection{Protonated formaldimine}
\label{res_pro_form_intro}

The photophysics of protonated formaldimine, \ce{CH2NH2+}, has been extensively studied (e.g., Refs.~\citenum{bonacic-koutecky_critically_1987, bonacic-koutecky_neutral_1987, du_theoretical_1990, el-taher_ab_2001, aquino_excited-state_2006}), with the molecule acting as the simplest protonated Schiff base model system for the rhodopsin chromophore, retinal. Within the FC region, protonated formaldimine possesses an optically dark S$_1$ state and a bright S$_2$ state of predominantly $\sigma\pi^*$ and $\pi\pi^*$ electronic character, respectively.\cite{barbatti_ultrafast_2006, tavernelli_non-adiabatic_2009} Given the much higher oscillator strength exhibited by S$_2$, photoexcitation occurs predominantly to S$_2$, with relaxation to the S$_0$ ground state involving passage through two subsequent MECXs. The first (S$_2$/S$_1$) has been shown to exhibit a peaked topography, whilst the second (S$_1$/S$_0$) has been shown to be sloped.\cite{barbatti_ultrafast_2006} Hence, protonated formaldimine constitutes a perfect model system (i.e., possessing MECXs (i) between different types of electronic states and (ii) exhibiting different topographies) to assess the quality of the branching space provided by (LR-TD)DFT/TDA/PBE0 and MP2/ADC(2).

\subsubsection{Linear interpolation in internal coordinates}
\label{res_pro_form_liic}

\begin{figure}[h!]
    \centering
    \includegraphics[width=\textwidth]{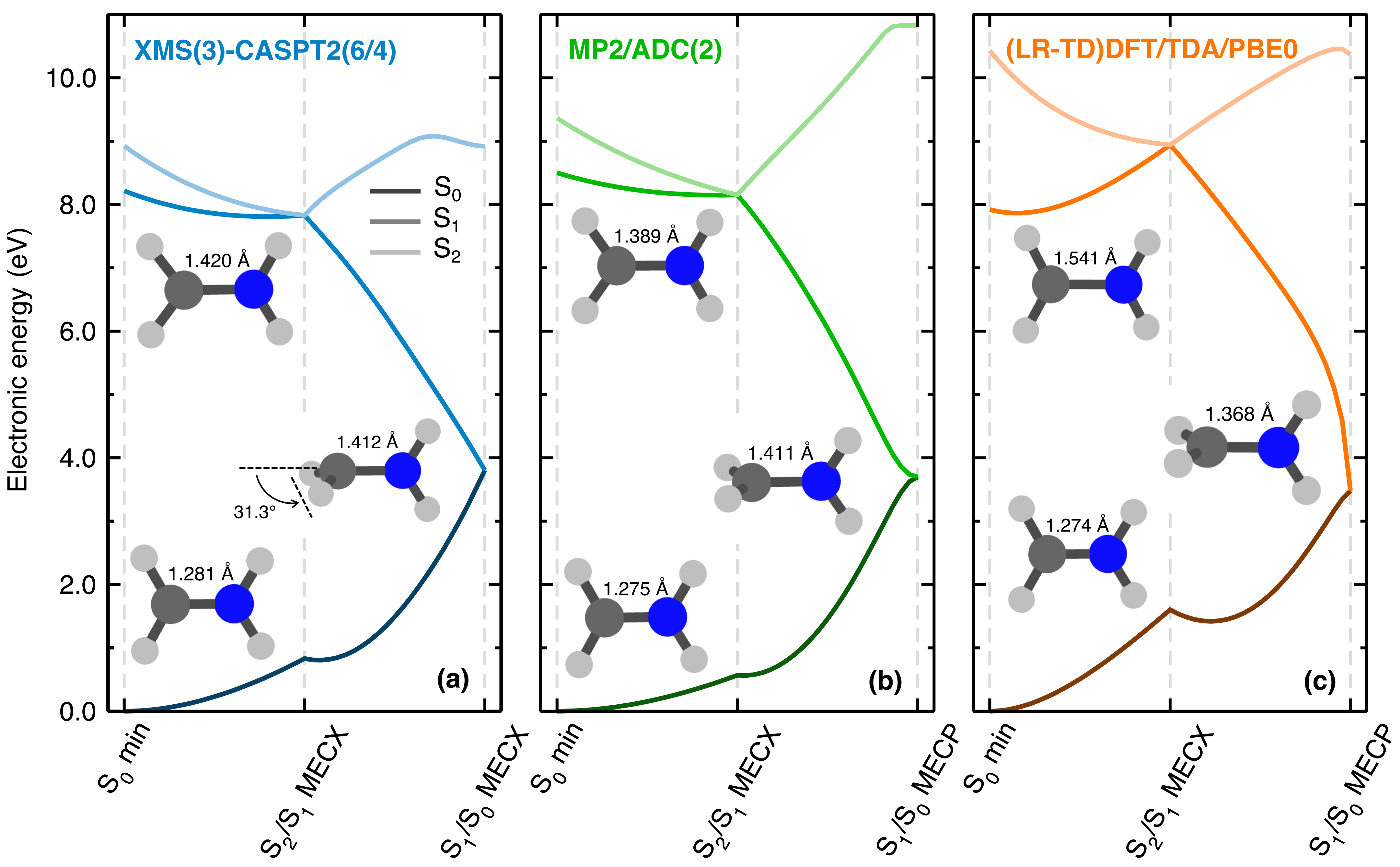}
    \caption{LIIC pathways connecting the S$_0$ minimum, S$_2$/S$_1$ MECX and S$_1$/S$_0$ MECX (or MECP) in protonated formaldimine. Comparison of the S$_0$ (dark colour), S$_1$ (mid colour) and S$_2$ (light colour) electronic energies obtained with (a) XMS(3)-CASPT2(6/4)/cc-pVTZ (blue), (b) MP2/ADC(2)/cc-pVTZ (green) and (c) (LR-TD)DFT/TDA/PBE0/cc-pVDZ (orange). In each panel, the critical geometries were optimised at the same level of theory used to compute the electronic energies. The insets show the molecular structures of the three critical points (S$_0$ min, bottom left; S$_2$/S$_1$ MECX, top left; S$_1$/S$_0$ MECX (or MECP), middle right) along with the calculated C-N bond lengths (and the CH$_2$ pyramidalisation angle for the XMS(3)-CASPT2(6/4) S$_1$/S$_0$ MECX geometry).}
    \label{fig:pro_form_liic}
\end{figure}

In the following, we compare the photochemical pathway of protonated formaldimine by calculating the three lowest electronic-state energies along an LIIC pathway connecting the FC, S$_2$/S$_1$ MECX and S$_1$/S$_0$ MECX critical geometries obtained with XMS-CASPT2, MP2/ADC(2) and (LR-TD)DFT/TDA/PBE0 (see molecular representation in Fig.~\ref{fig:pro_form_liic}).

According to XMS-CASPT2 (Fig.~\ref{fig:pro_form_liic}a), following photoexcitation to S$_2$, protonated formaldimine decays to S$_1$ via a strongly peaked S$_2$/S$_1$ MECX, which is encountered by a stretch of the C-N bond whilst retaining the planarity of the molecule exhibited at the FC geometry (S$_0$ min, 1.281 \AA; S$_2$/S$_1$ MECX, 1.420 \AA). Such a peaked topography is assumed to provide highly efficient nonadiabatic population transfer from S$_2$ to S$_1$.\cite{barbatti_ultrafast_2006} Subsequent relaxation to the ground electronic state occurs through a weakly sloped S$_1$/S$_0$ MECX, which for XMS-CASPT2 is reached via a 90$^\circ$ twist of the C-N bond and an additional 31.3$^\circ$ pyramidalisation of the CH$_2$ moiety. Less efficient S$_1$-to-S$_0$ decay is expected for the predicted sloped topography of the S$_1$/S$_0$ MECX.\cite{barbatti_ultrafast_2006} Previous investigations with MRCISD have reported purely twisted S$_1$/S$_0$ MECX geometries with no CH$_2$ pyramidalisation,\cite{levine_optimizing_2008, nikiforov_assessment_2014} whereas others employing MS-CASPT2 have instead predicted the C-N torsion accompanied by pyramidalisation of the NH$_2$ group.\cite{levine_optimizing_2008} The differences in S$_1$/S$_0$ MECX geometry obtained by different multireference methods have been ascribed to an apparent flatness of the intersection seam with respect to pyramidalisation (at either end of the C-N bond).\cite{levine_optimizing_2008}

We now compare the XMS-CASPT2 LIIC pathway to those obtained with (LR-TD)DFT/TDA/PBE0 (Fig.~\ref{fig:pro_form_liic}c) and MP2/ADC(2) (Fig.~\ref{fig:pro_form_liic}b). Considering the overall electronic energy profiles of the different methods along the LIIC, an obvious observation is the striking agreement between MP2/ADC(2) and XMS-CASPT2; the only notable difference is the behaviour of S$_2$ in the segment connecting the two MECXs (explained by the involvement of other electronic states not included in XMS-CASPT2). On the other hand, LR-TDDFT/TDA/PBE0 predicts an ${\rm S}_2 - {\rm S}_1$ energy difference at the S$_0$ minimum over twice that given by either ADC(2) or XMS-CASPT2. The approach to the respective MECX (or MECP) points are also markedly different in (LR-TD)DFT/TDA/PBE0 compared to that in the wavefunction-based methods. Notably, the LR-TDDFT/TDA/PBE0 S$_1$ state approaches the S$_1$/S$_0$ MECP too steeply relative to XMS-CASPT2. This observation further corroborates that the LR-TDDFT/TDA first excited electronic state can vary too rapidly in the vicinity of a CX with the ground state, as previously shown in Ref. \onlinecite{levine_conical_2006}. Interestingly, neither MP2/ADC(2) nor (LR-TD)DFT/TDA/PBE0 predict the CH$_2$ pyramidalisation exhibited by XMS-CASPT2 for the S$_1$/S$_0$ MECX geometry, despite all three geometries being at approximately the same relative energy. Earlier works using (LR-TD)DFT/TDA/PBE presented similar observations.\cite{tavernelli_non-adiabatic_2009} 

\subsubsection{S$_2$/S$_1$ branching space}
\label{res_pro_form_s2s1_branch_space}

\begin{figure}[h!]
    \centering
    \includegraphics[width=0.45\textwidth]{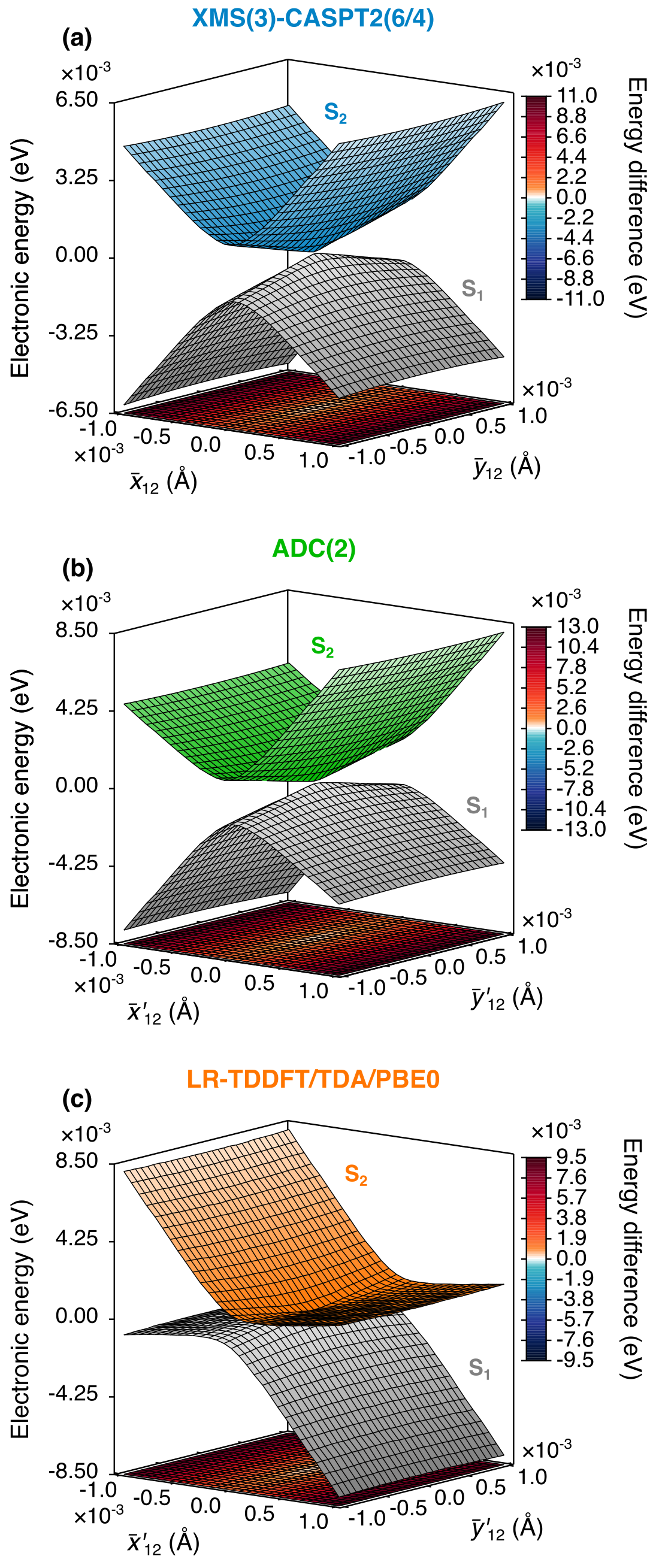}
    \caption{Branching space of the S$_2$/S$_1$ MECX in protonated formaldimine. Comparison of the S$_1$ and S$_2$ PESs obtained with (a) XMS(3)-CASPT2(6/4)/cc-pVTZ (blue/grey), (b) ADC(2)/cc-pVTZ (green/grey) and (c) LR-TDDFT/TDA/PBE0/cc-pVDZ (orange/grey). In each plot, the MECX geometries and branching space vectors were obtained at the same level of theory used to calculate the electronic energies (except for the ADC(2) plot, which used the $\h{1}{2}$ vector of XMS(3)-CASPT2(6/4) -- see Sec.~\ref{comp_dets_branch_space} for details). The base in each plot shows a 2D colour map of the ${\rm S}_2 - {\rm S}_1$ energy difference (see colour bar on the right).}
    \label{fig:pro_form_s2s1_bp}
\end{figure}

We now focus our attention on the first intersection seam encountered by protonated formaldimine upon photoexcitation to S$_2$, by calculating the electronic energies with each electronic structure method within the branching space of their respective S$_2$/S$_1$ MECX (Fig.~\ref{fig:pro_form_s2s1_bp}). All three electronic structure methods correctly predict a conical ($F - 2$)-dimensional intersection between S$_1$ and S$_2$, where the degeneracy is lifted in both branching space vector directions. We stress again (see Section \ref{tddft_vs_cxs}) that the success of LR-TDDFT/TDA to accurately describe the topology of the S$_2$/S$_1$ MECX is \textit{not necessarily guaranteed}. Our results, however, confirm that linear-response $\h{1}{2}$ vectors do indeed offer an adequate description of the CX branching space in protonated formaldimine.

We note that the S$_1$ and S$_2$ PESs obtained with LR-TDDFT/TDA/PBE0 are in relatively poor agreement with those of the XMS-CASPT2 reference (compare Fig.~\ref{fig:pro_form_s2s1_bp}c to Fig.~\ref{fig:pro_form_s2s1_bp}a). Using the CX branching space topography parameters defined in Ref.~\onlinecite{fdez_galvan_analytical_2016}, both methods yield a peaked bifurcating topography, but LR-TDDFT/TDA/PBE0 exhibits larger values of $\mathcal{P}$ and $\mathcal{B}$ (0.59 and 0.86, respectively) than XMS-CASPT2 (0.02 and 0.29). These parameters are summarised in the SI. This disparity between LR-TDDFT/TDA/PBE0 and XMS-CASPT2 links to the LIIC plots in Fig.~\ref{fig:pro_form_liic}, where the approach of the LR-TDDFT/TDA/PBE0 S$_2$ and S$_1$ states (i.e., the ${\rm S}_2 - {\rm S}_1$ energy gap and slope of the S$_2$ and S$_1$ energies) towards the S$_2$/S$_1$ MECX is markedly different in LR-TDDFT/TDA/PBE0 to that in either XMS-CASPT2 or ADC(2). On the other hand, the S$_1$ and S$_2$ PESs obtained with ADC(2) are in close agreement with those of XMS-CASPT2; ADC(2) also yields a peaked bifurcating topography for the S$_2$/S$_1$ MECX (compare Fig.~\ref{fig:pro_form_s2s1_bp}b to Fig.~\ref{fig:pro_form_s2s1_bp}a) with similar parameter values of $\mathcal{P}=0.08$ and $\mathcal{B}=0.45$. The ability of ADC(2) to adequately describe the branching space of a CX between excited electronic states is reassuring, given its extended use within excited-state molecular dynamics simulations.\cite{plasser_surface_2014, bennett_nonadiabatic_2016, kochman_theoretical_2019, sapunar_timescales_2015, prlj_excited_2015, kochman_simulating_2020, siddique_nonadiabatic_2020, barbatti_photorelaxation_2014, lischka_effect_2018, milovanovic_simulation_2021, prlj_rationalizing_2016, novak_photochemistry_2017, dommett_excited_2017, gate_photodynamics_2019, kochman_theoretical_2016, hutton_photodynamics_2022, szabla_ultrafast_2016} 

\subsubsection{S$_1$/S$_0$ branching space}
\label{res_pro_form_s1s0_branch_space}

\begin{figure}[h!]
    \centering
    \includegraphics[width=0.41\textwidth]{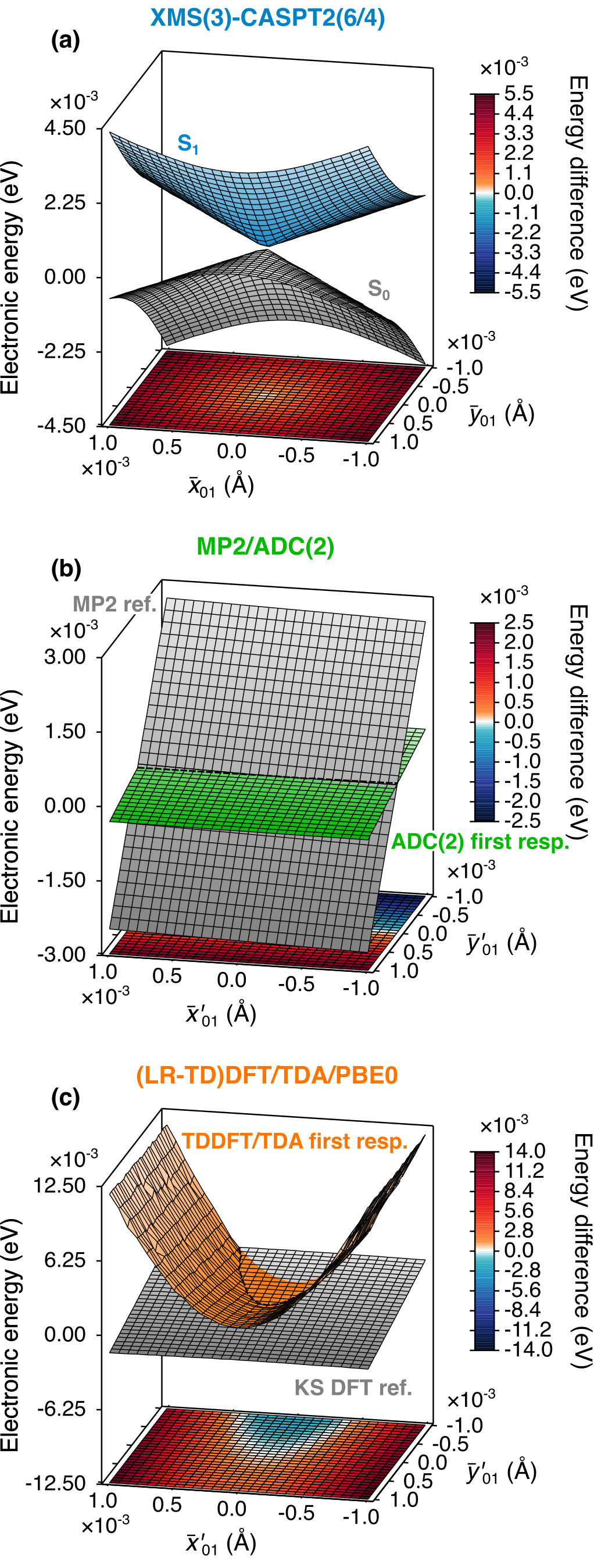}
    \caption{Branching space of the S$_1$/S$_0$ MECX (or MECP) in protonated formaldimine. Comparison of the S$_0$ and S$_1$ PESs obtained with (a) XMS(3)-CASPT2(6/4)/cc-pVTZ (blue/grey), (b) MP2/ADC(2)/cc-pVTZ (green/grey) and (c) (LR-TD)DFT/TDA/PBE0/cc-pVDZ (orange/grey). In each plot, the MECX (or MECP) geometries and branching space vectors were obtained at the same level of theory used to calculate the electronic energies (except for the MP2/ADC(2) plot, which used the $\h{0}{1}$ vector of XMS(3)-CASPT2(6/4) -- see Sec.~\ref{comp_dets_branch_space} for details). The dashed lines in the Figs. 3(b) and 3(c) indicate the seam where $E_{0}(\R) = E_{1}(\R)$. (We note that the rendering of the colours for the PESs does not reflect precisely this intersection.) The base in each plot shows a 2D colour map of the ${\rm S}_1 - {\rm S}_0$ energy difference (see colour bar on the right).}
    \label{fig:pro_form_s1s0_bp}
\end{figure}

Next, we take the opportunity to focus on the performance of the methods in describing the S$_1$/S$_0$ MECX branching space of protonated formaldimine. XMS-CASPT2 gives a conical ($F - 2$)-dimensional intersection as expected (Fig.~\ref{fig:pro_form_s1s0_bp}a), with a sloped single-path topography (with parameters, $\mathcal{P}=1.49$ and $\mathcal{B}=1.32$) similar to that reported in Ref.~\citenum{barbatti_ultrafast_2006}. As expected from the discussion in Sec.~\ref{tddft_vs_cxs}, (LR-TD)DFT/TDA/PBE0 and MP2/ADC(2) incorrectly predict a linear ($F - 1$)-dimensional intersection at the S$_1$/S$_0$ MECP (Fig.~\ref{fig:pro_form_s1s0_bp}c and Fig.~\ref{fig:pro_form_s1s0_bp}b, respectively), where the degeneracy is only lifted along a single branching space vector direction (i.e., $\ya{0}{1}$). In both cases, the first response (S$_1$) state becomes lower in energy than the reference (S$_0$) state, leading to negative excitation energies along certain regions of the branching plane (see colourmap in Fig.~\ref{fig:pro_form_s1s0_bp}b and Fig.~\ref{fig:pro_form_s1s0_bp}c). This observation corroborates earlier results obtained for (LR-TD)DFT\cite{levine_conical_2006} and MP2/ADC(2).\cite{tuna_assessment_2015} When plotted using the same vertical axis energy range (see Fig. S2 in the SI), it is clear that the (LR-TD)DFT/TDA/PBE0 S$_1$ PES varies too rapidly in the vicinity of the S$_1$/S$_0$ MECP compared to that of both XMS-CASPT2 (where a conical intersection is obtained), and MP2/ADC(2) (where a linear seam of intersection is observed). This difference in behaviour between the different electronic structure methods is consistent with the LIIC plots in Fig.~\ref{fig:pro_form_liic} close to the S$_1$/S$_0$ intersection region. (We note that replacing the (LR-TD)DFT branching space vectors used to generate the (LR-TD)DFT S$_1$/S$_0$ MECP (and S$_2$/S$_1$ MECX) branching space plots in Fig.~\ref{fig:pro_form_s1s0_bp} (and \ref{fig:pro_form_s2s1_bp}) with those of XMS-CASPT2 results in no observable difference to the PESs -- except for a trivial reflection in the $\ya{i}{j}$ vector direction.)

\begin{figure}[h!]
    \centering
    \includegraphics[width=0.65\textwidth]{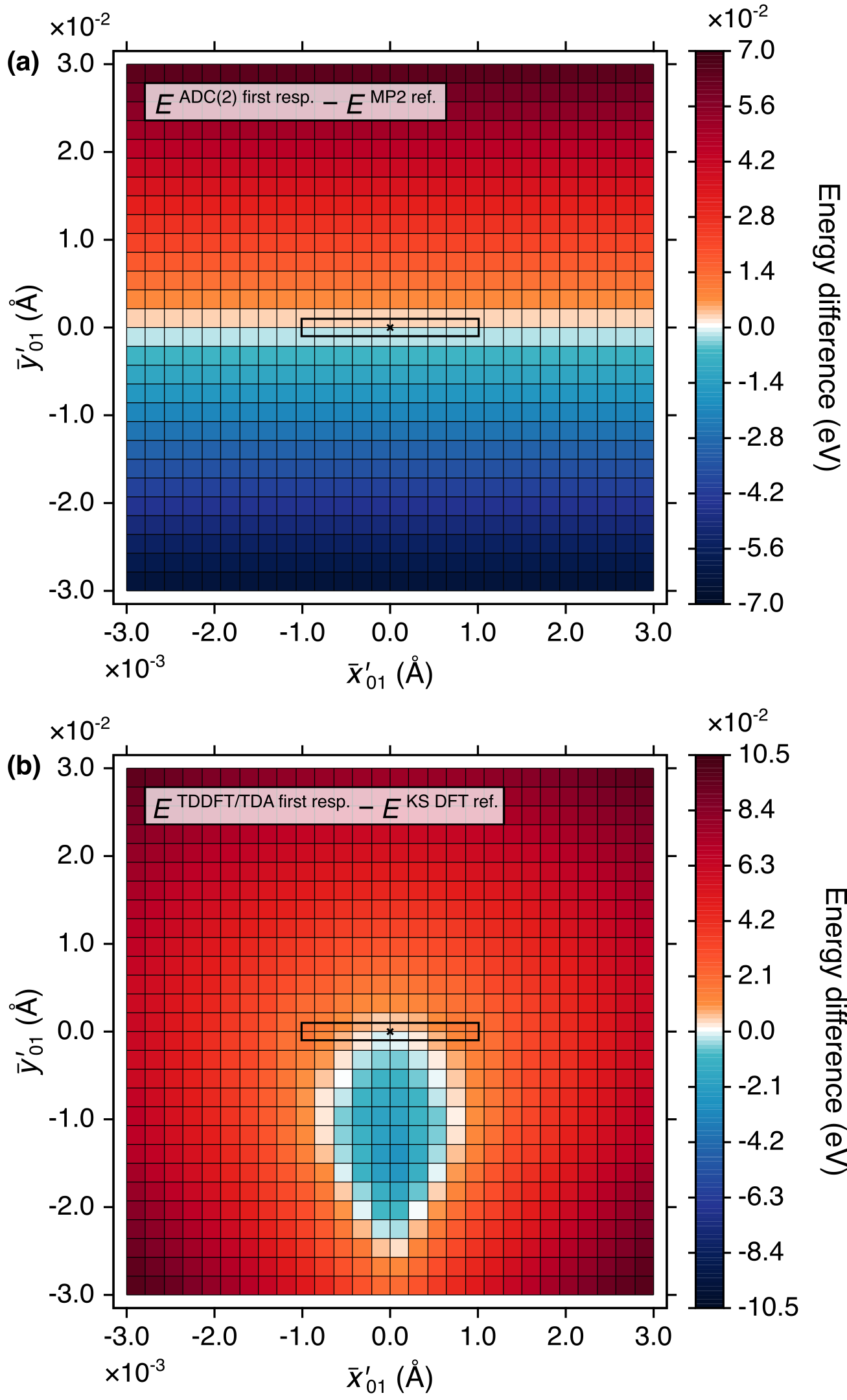}
    \caption{2D colour map of the electronic energy difference between S$_0$ (reference state) and S$_1$ (first response state) obtained with (a) MP2/ADC(2)/cc-pVTZ and (b) (LR-TD)DFT/TDA/PBE0/cc-pVDZ in the vicinity of the S$_1$/S$_0$ MECP along an extended branching plane ($\pm 0.003 \times \xa{0}{1}$, $\pm 0.03 \times \ya{0}{1}$). The black box encloses the area spanned by the branching plane used to generate the plots in Fig.~\ref{fig:pro_form_s1s0_bp}; the black cross indicates the location of the MECP geometry.}
    \label{fig:pro_form_s1s0_zoom_out}
\end{figure}

Despite indeed being ($F - 1$)-dimensional near the point where the two electronic states become degenerate, the (LR-TD)DFT/TDA/PBE0 intersection in Fig.~\ref{fig:pro_form_s1s0_bp}c appears significantly more curved than the strictly linear S$_1$/S$_0$ intersection of MP2/ADC(2) in Fig.~\ref{fig:pro_form_s1s0_bp}b. This observation warrants further investigation of the (LR-TD)DFT/TDA/PBE0 intersection at larger distances along the $\ya{0}{1}$ vector direction. Plotting the (LR-TD)DFT/TDA/PBE0 S$_0$ and S$_1$ PESs along an extended branching plane ($\pm 0.003 \times \xa{0}{1}$ and $\pm 0.03 \times \ya{0}{1}$ in Fig.~\ref{fig:pro_form_s1s0_zoom_out}b compared to the original $\pm 0.001 \times \xa{0}{1}$ and $\pm 0.001 \times \ya{0}{1}$ in Fig.~\ref{fig:pro_form_s1s0_bp} -- see SI for branching space vector definitions) reveals that the curved intersection seam in Fig.~\ref{fig:pro_form_s1s0_bp}c is in fact just one part of a larger intersection ring - something that shows a striking resemblance to two interpenetrating cones. On the other hand, the strictly linear intersection seam in MP2/ADC(2) observed along the standard branching plane (Fig.~\ref{fig:pro_form_s1s0_bp}b) remains even along this extended branching plane (Fig.~\ref{fig:pro_form_s1s0_zoom_out}a). Overall, our results connect the different pictures proposed earlier for the description of S$_1$/S$_0$ MECPs within (LR-TD)DFT/TDA: performing an S$_1$/S$_0$ MECP optimisation with (LR-TD)DFT/TDA will in fact locate a geometry on the intersection ring and the MECP will look different depending on the extent of the branching space explored to unravel the shape of the S$_0$ and  S$_1$ PESs around this location -- either a (near-to-linear) seam of intersection for minute variations along $\xa{0}{1}$ and $\ya{0}{1}$ (like in Fig.~\ref{fig:pro_form_s1s0_bp}c and as first reported by Levine \textit{et al.}\cite{levine_conical_2006}), or an \textit{intersection ring} (reminiscent of two interpenetrating cones) when a more extended scan along $\xa{0}{1}$ and $\ya{0}{1}$ is performed (like in Fig.~\ref{fig:pro_form_s1s0_zoom_out}b and as alluded to by Tapavicza \textit{et al.}\cite{tapavicza_mixed_2008}). We note that it may be possible to miss the negative-energy region of the intersection ring for more extreme scans around the (LR-TD)DFT/TDA S$_1$/S$_0$ intersection point (i.e., if one "zooms out" further from the crossing point), giving a false impression that (LR-TD)DFT/TDA can describe the intersection point adequately.

We conclude this Section by noting that we also calculated the HF/CIS S$_1$/S$_0$ MECP branching space for both the standard and extended grid of geometries around the intersection point -- see Figs. S4 and S5 in the SI. As expected (Section \ref{tddft_vs_cxs}) HF/CIS predicts a strictly linear ($F - 1$)-dimensional intersection along the standard branching plane that likewise remains along the extended branching plane, which is analogous to the behaviour of MP2/ADC(2), but in contrast to that of (LR-TD)DFT/TDA/PBE0. We have confirmed that our (LR-TD)DFT/TDA findings are unaffected by improving the numerical accuracy of our calculations (i.e., increased grid size -- see also the SI for details regarding SCF convergence). These observations solidify our conclusions that the description of CXs involving the ground state by (LR-TD)DFT/TDA and HF/CIS are not completely analogous. 

\subsection{Pyrazine}

\begin{figure}[h!]
    \centering
    \includegraphics[width=0.45\textwidth]{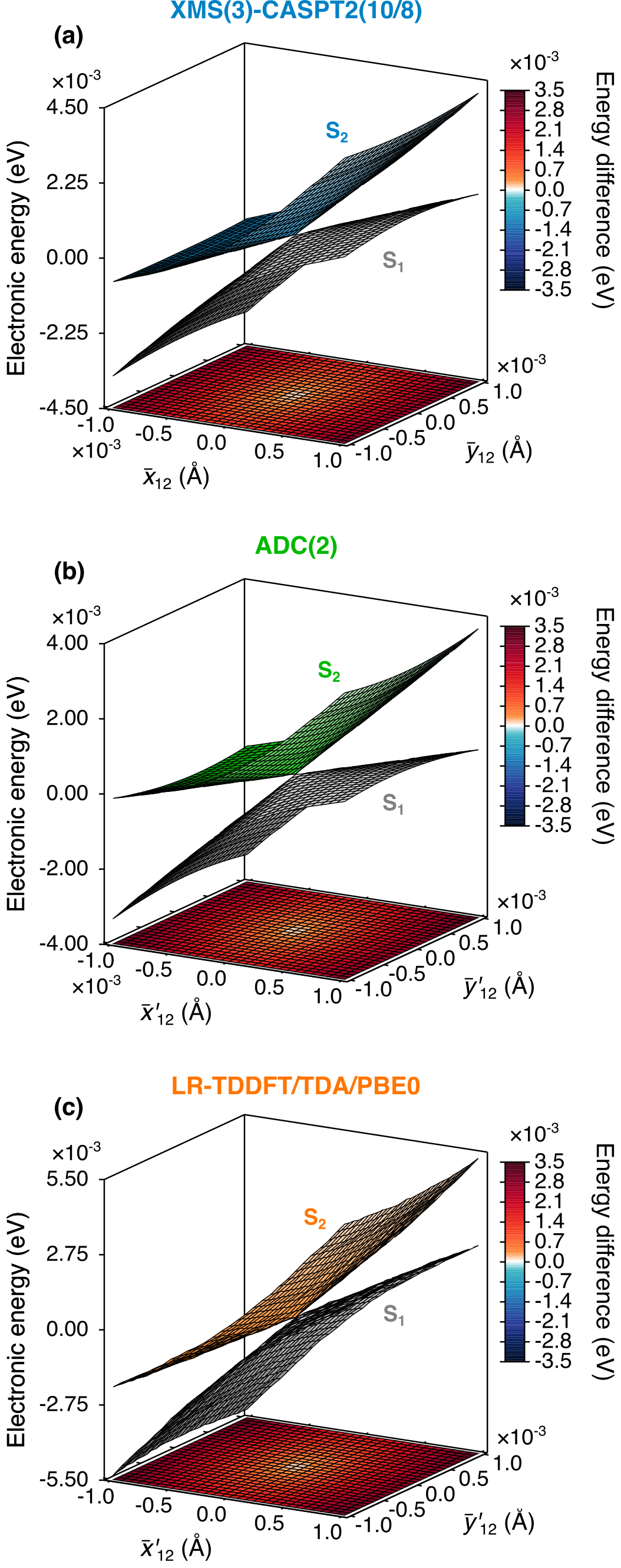}
    \caption{Branching space of the S$_2$/S$_1$ MECX in pyrazine. Comparison of the S$_1$ and S$_2$ PESs obtained with (a) XMS(3)-CASPT2(6/4)/cc-pVTZ (blue/grey), (b) ADC(2)/cc-pVTZ (green/grey) and (c) LR-TDDFT/TDA/PBE0/cc-pVDZ (orange/grey). In each plot, the MECX geometries and branching space vectors were obtained at the same level of theory used to calculate the electronic energies (except for the ADC(2) plot, which used the $\h{1}{2}$ vector of XMS-CASPT2(10/8) -- see Sec.~\ref{comp_dets_branch_space} for details). The base in each plot shows a 2D colour map of the ${\rm S}_2 - {\rm S}_1$ energy difference (see colour bar on the right).}
    \label{fig:pyra_s2s1_bp}
\end{figure}

Next, we consider CXs between excited states for a second exemplar molecule, pyrazine. Like for protonated formaldimine, the excited electronic states of pyrazine have been well-studied, often considered the definitive case for vibronic coupling in aromatic systems; pyrazine is also a precursor to numerous biologically active molecules.\cite{woywod_theoretical_2010, stock_resonance_1995, sala_quantum_2015, shiozaki_pyrazine_2013, woywod_characterization_1994, kanno_ab_2015, sobolewski_ab_1993, seidner__1992} Within the FC region, the S$_1$ state in pyrazine exhibits an $n\pi^*$ character and S$_2$ is of $\pi\pi^*$ character.\cite{woywod_characterization_1994} At the XMS-CASPT2 level, the S$_2$/S$_1$ MECX is reached (from the planar S$_0$ minimum geometry) by simultaneous elongation of the C-N and C-C bonds, but with an overall stretching of the ring along the axis bisecting the two nitrogen atoms (see Fig. S6 in the SI). LR-TDDFT/TDA/PBE0 and ADC(2) predict S$_0$ minimum and S$_2$/S$_1$ MECX geometries that agree closely with those of XMS-CASPT2. The only difference is the stretching of the S$_2$/S$_1$ MECX geometry observed in LR-TDDFT/TDA/PBE0 is slightly more exaggerated than in the wavefunction-based methods, as indicated by the larger (smaller) N-C-C (C-N-C) bond angles. This distortion in the LR-TDDFT/TDA/PBE0 S$_2$/S$_1$ MECX geometries is accompanied by it being approximately 1 eV higher in energy than the S$_2$/S$_1$ MECX geometry in either XMS-CASPT2 or ADC(2).

\subsubsection{S$_2$/S$_1$ branching space}
\label{res_pyra_s2s1_branch_space}

We focus on the respective branching spaces for the S$_2$/S$_1$ MECX (Fig.~\ref{fig:pyra_s2s1_bp}). As for protonated formaldimine, all three methods correctly predict a conical ($F - 2$)-dimensional intersection between S$_1$ and S$_2$, where the degeneracy is lifted along both branching space vector directions. LR-TDDFT/TDA/PBE0 exhibits a sloped single-path MECX, mirroring the topography observed with both XMS-CASPT2 and ADC(2), with $\mathcal{P}$ and $\mathcal{B}$ parameters (7.16 and 2.64, respectively) that are closer to those obtained with XMS-CASPT2 (3.57 and 1.96) than ADC(2) (12.78 and 1.14). Recalculating the S$_2$/S$_1$ MECX geometry and its corresponding branching space with a different exchange-correlation functional (i.e., the range-separated hybrid LC-$\omega$PBE, with range-separated parameter $\omega$ = 0.4 $a_0^{-1}$ -- see Fig. S7 in the SI) further generalises our findings and conclusions that LR-TDDFT/TDA can adequately reproduce the dimensionality of a CX between excited electronic states.

\section{Conclusion}
\label{conc}

This work has shown explicitly that LR-TDDFT/TDA/PBE0 within the AA is able to exhibit the correct topology of a CX between two \textit{excited} electronic states for two exemplar molecules, protonated formaldimine and pyrazine. The correct CX topology was unchanged when an alternative exchange-correlation functional was investigated for pyrazine. We further showed that ADC(2) offers an accurate description of both the topology and topography of CXs between excited electronic states, and note that this is in contrast to that of (conventional) coupled cluster theory, which can be flawed in this context.\cite{plasser_surface_2014, hattig_structure_2005, kohn_can_2007, kjonstad_crossing_2017, thomas_complex_2021} We stress that all CX branching spaces analysed in this work were constructed within a fully-consistent approach where all required electronic quantities were computed (where possible) at the same level of theory. 

Re-inspection of the problem faced by AA (LR-TD)DFT/TDA to adequately describe CXs involving the ground electronic states also proved fruitful. Our findings for protontated formaldimine show that the two, supposedly different, pictures related to the S$_1$/S$_0$ MECP branching space of AA (LR-TD)DFT/TDA/PBE0 -- a seam of intersection vs. two interpenetrating cones -- both emanate from the intersection ring, which can be reconciled by analysing the behaviour of the PESs, either in the immediate vicinity of the S$_1$/S$_0$ MECP, or at further distances from the MECP geometry. The intersection ring from AA (LR-TD)DFT/TDA/PBE0 is in stark contrast to the linear intersection observed from MP2/ADC(2) (and, as expected, HF/CIS). Further work is arguably still needed to pinpoint precisely how nonadiabatic dynamics simulations is influenced by the intersection ring and whether the difference in behaviour of AA (LR-TD)DFT/TDA/PBE0 to that of HF/CIS gives any grounds for optimism when applying AA (LR-TD)DFT/TDA in this context. Again, extending the use of previously proposed expressions for the exact, frequency-dependent exchange-correlation kernel\cite{maitra_double_2004, casida_propagator_2005, ferre_many-body_2015, romaniello_double_2009, gritsenko_double_2009, strinati_application_1988, zhang_dynamical_2013, authier_dynamical_2020, thiele_frequency_2014} to the problem of CXs involving the ground electronic state still remains as pertinent as ever. Nonetheless, for the case of CXs between excited electronic states, greater confidence (at least for electronic states dominated by single excitations) should be felt when applying AA LR-TDDFT/TDA to chemically (and biologically) relevant systems, whose size still prohibits the use of multiconfigurational methods.

\section*{Acknowledgements}

The authors thank Prof. Neepa T. Maitra, Prof. E. K. U. Gross, and Prof. Todd J. Mart\'{i}nez for stimulating discussions. This project has received funding from the European Research Council (ERC) under the European Union's Horizon 2020 research and innovation programme (Grant agreement No. 803718, project SINDAM) and the EPSRC Grant EP/V026690/1. JTT acknowledges the EPSRC for an EPSRC Doctoral Studentship (EP/T518001/1). 
This work made use of the facilities of the Hamilton HPC Service of Durham University.

\section*{Data availability}

The data that supports the findings of this study are available within the article and its supplementary material.


\end{document}


\newcommand\bfec[1]{\textcolor{blue}{\textit{ XXX BFEC: #1 XXX }}}
\newcommand\jtt[1]{\textcolor{orange}{\textit{ XXX JTT: #1 XXX }}}
\newcommand\djt[1]{\textcolor{green}{\textit{ XXX DJT: #1 XXX }}}

\newcommand{\g}[2]{\mathbf{g}_{#1#2}}
\newcommand{\h}[2]{\mathbf{h}_{#1#2}}
\newcommand{\nacv}[2]{\mathbf{d}_{#1#2}}
\newcommand{\R}{\mathbf{R}}
\renewcommand{\thefigure}{S\arabic{figure}}
\renewcommand{\thetable}{S\arabic{table}}
\setlength{\tabcolsep}{12pt}

\DeclareFontFamily{OT1}{cmbr}{\hyphenchar\font45 }
\DeclareFontShape{OT1}{cmbr}{m}{n}{%
  <-9>cmbr8
  <9-10>cmbr9
  <10-17>cmbr10
  <17->cmbr17
}{}
\DeclareFontShape{OT1}{cmbr}{m}{sl}{%
  <-9>cmbrsl8
  <9-10>cmbrsl9
  <10-17>cmbrsl10
  <17->cmbrsl17
}{}
\DeclareFontShape{OT1}{cmbr}{m}{it}{%
  <->ssub*cmbr/m/sl
}{}
\DeclareFontShape{OT1}{cmbr}{b}{n}{%
  <->ssub*cmbr/bx/n
}{}
\DeclareFontShape{OT1}{cmbr}{bx}{n}{%
  <->cmbrbx10
}{}

\title{Supporting information: \\ On the description of conical intersections between excited electronic states with LR-TDDFT and ADC(2)}
\author{Jack T. Taylor}
\author{David J. Tozer}
\email{d.j.tozer@durham.ac.uk}
\affiliation{Department of Chemistry, Durham University, South Road, Durham DH1 3LE, UK}
\author{Basile F. E. Curchod}
\email{basile.curchod@bristol.ac.uk}
\affiliation{Centre for Computational Chemistry, School of Chemistry, University of Bristol, Cantock's Close, Bristol BS8 1TS, UK}
\date{\today}
\begin{abstract}
    \tableofcontents
\end{abstract}

\maketitle

\section{Supplementary computational details}

\subsection{Electronic structure}

All electronic energies, nuclear gradients, and nonadiabatic coupling vectors that were computed with the range-separated LC-$\omega$PBE functional, HF and CIS\cite{szabo_modern_1996} were obtained using a development version of the GPU-accelerated TeraChem 1.9 program.\cite{isborn_excited-state_2011, ufimtsev_quantum_2008, ufimtsev_quantum_2009, ufimtsev_quantum_2009-1, titov_generating_2013, seritan_terachem_2020, seritan_span_2021} In LC-$\omega$PBE, a range-separation parameter of 0.4 Bohr$^{-1}$ was used and all AA LR-TDDFT calculations were performed within the Tamm-Dancoff approximation.\cite{hirata_time-dependent_1999} The Dunning cc-pVDZ basis set was used in all DFT, AA LR-TDDFT, HF and CIS calculations.\cite{dunning_gaussian_1989}

\subsection{Critical points and linear interpolation in internal coordinates}

An iterative procedure was used to locate the MECXs (or MECPs) in (LR-TD)DFT/TDA using TeraChem. Each iteration involved performing four separate MECX geometry optimisation calculations using four different algorithms: (i) the gradient-projection method of Bearpark \textit{et al.},\cite{bearpark_direct_1994} (ii) the Lagrange-Newton method of Manaa and Yarkony,\cite{manaa_intersection_1993} (iii) the penalty-function method of Ciminelli \textit{et al.}\cite{ciminelli_photoisomerization_2004} and (iv) the CIOpt approach (with a CIGap criterion of 0.001 a.u.) of Levine \textit{et al.}\cite{levine_optimizing_2008} Methods (i)-(iii) were used within TeraChem, whilst method (iv) was externally interfaced to TeraChem. The input geometry for the four MECX optimisation calculations of a given iteration of the procedure was the geometry obtained from the previous iteration that best satisfied the following criteria: the geometry (a) possesses the smallest $E_j(\R) - E_i(\R)$ energy gap (i.e., represents a CX (or CP)), (b) possesses the lowest $E_i(\R)$ and $E_j(\R)$ energies (i.e., represents a minimum-energy CX (or CP)) and (c) is chemically sensible/relevant (i.e., corresponds to the MECX (or MECP) of interest). The XMS-CASPT2/cc-pVTZ geometries were used as the initial input geometries for the first iteration of the procedure. The process was repeated until either, all four MECX optimisation algorithms converged on the same geometry, or until the current iteration resulted in an MECX (or MECP) geometry that satisfied the above criteria less well than the geometry obtained in the previous iteration. The procedure took 3-4 iterations in all cases. The final geometry was then taken to be \textit{the} optimised (LR-TD)DFT/TDA MECX (or MECP) geometry.

Similarly, an iterative procedure was used to optimise the MECXs (or MECPs) with MP2/ADC(2). Four consecutive MECX optimisation calculations were performed using CIOpt\cite{levine_optimizing_2008} externally interfaced to Turbomole, where the geometry obtained in the previous optimisation served as the input to the current optimisation. The CIGap criterion was set to 0.01 a.u. for the first MECX optimisation and was decreased by an order of magnitude for each subsequent optimisation, with the XMS-CASPT2/cc-pVTZ geometries again being used as the initial input geometry. After the four consecutive MECX optimisation calculations, the geometry that best satisfied the three criteria (i) to (iii) in the above paragraph was taken to be the preliminary MP2/ADC(2) MECX (or MECP) geometry.  Further refinement was performed by computing the raw branching space vectors at this geometry and generating its corresponding raw branching space. The branching space geometry with the smallest $E_j(\R) - E_i(\R)$ energy gap (if different to the original MECX -- or MECP -- geometry located at the origin) was then redefined as \textit{the} optimised MP2/ADC(2) MECX (or MECP) geometry.

The same procedure used to locate the MECP in (LR-TD)DFT/TDA was used to optimise the MECP geometry in HF/CIS. Further refinement of the HF/CIS S$_1$/S$_0$ MECP geometry was carried out by generating the corresponding raw branching space, analogous to the procedure used in optimising the MECX (or MECP) geometries in MP2/ADC(2).

\subsection{Plotting the CX branching space}

\subsubsection{Orthonormalisation (and alignment) of branching space vectors}

To generate the branching space vectors used to construct the MECX (or MECP) branching space plots investigated in this work, we followed the approach taken in Ref. \onlinecite{liu_analytical_2021}, which we summarise here. The raw branching space vectors, $\g{i}{j}(\R)$ and $\h{i}{j}(\R)$, were first computed directly at the optimised S$_j$/S$_i$ MECX (or MECP) geometry. (Note, the $\h{i}{j}(\R)$ vectors were first obtained from the $\nacv{i}{j}(\R)$ vectors via the following relationship: $\h{i}{j}(\R) = [E_j(\R) - E_i(\R)] \times \nacv{i}{j}(\R)$. For clarity, we will drop the explicit dependence on '$\R$' hereafter.) Raw branching space vectors computed using a given electronic structure method (at an MECX - or MECP - geometry optimised using finite numerical accuracy) are, in general, not orthogonal with respect to each other.\cite{ferre_description_2015} As such, we used the Yarkony procedure\cite{yarkony_adiabatic_2000, yarkony_conical_2001} to obtain the following orthogonalised versions of the branching space vectors,
\begin{equation}
    \begin{split}
        \bar{\mathbf{g}}_{ij} &= \mathbf{g}_{ij}{\rm cos}(2\beta) + \mathbf{h}_{ij}{\rm sin}(2\beta) \,\,\, ,\\
        \bar{\mathbf{h}}_{ij} &= \mathbf{h}_{ij}{\rm cos}(2\beta) - \mathbf{g}_{ij}{\rm sin}(2\beta) \,\,\, .
    \end{split}
\end{equation}
To ensure orthogonality between $\bar{\mathbf{g}}_{ij}$ and $\bar{\mathbf{h}}_{ij}$, the required value of $\beta$ is determined by satisfying the following equation, 
\begin{equation}
    \begin{split}
        {\rm tan}(4\beta) = \frac{2 \left(\mathbf{g}_{ij} \cdot \mathbf{h}_{ij}\right)}{\left(\mathbf{g}_{ij} \cdot \mathbf{g}_{ij}\right) - \left(\mathbf{h}_{ij} \cdot \mathbf{h}_{ij}\right)} \,\,\, ,
    \end{split}
    \label{eq:beta}
\end{equation}
where $\beta = \beta_0 + n\frac{\pi}{4}$ for $n = 0, \pm1, \pm 2, ...$ and $\beta_0$ is the initial value solution of Eq.~\eqref{eq:beta} as stated in Ref. \onlinecite{liu_analytical_2021}. (Note, this orthogonalisation procedure does not change the branching plane obtained by a given electronic structure method, but simply orthogonalises the branching space vectors within the plane.)\cite{ferre_description_2015} The $\bar{\mathbf{g}}_{ij}$ and $\bar{\mathbf{h}}_{ij}$ vectors were then normalised to give the corresponding unit vectors (or intersection-adapted coordinates), 
\begin{equation}
    \bar{\mathbf{x}}_{ij} = \frac{\bar{\mathbf{g}}_{ij}}{||\bar{\mathbf{g}}_{ij}||} \hspace{0.25in} {\rm and} \hspace{0.25in} \bar{\mathbf{y}}_{ij} = \frac{\bar{\mathbf{h}}_{ij}}{||\bar{\mathbf{h}}_{ij}||} \,\,\, .
    \label{eq:unalign_vecs}
\end{equation}
These orthonormalised branching space vectors are uniquely defined (up to trivial transpositions and changes of sign).\cite{fdez_galvan_analytical_2016, liu_analytical_2021}

To facilitate direct comparison between the MECX (or MECP) branching space plots of the electronic structure method of \textit{interest} and that of the \textit{reference} electronic structure method, XMS-CASPT2, we carried out the following. We rotated the orthonormalised branching space vectors of the electronic structure method of interest, now given as
\begin{equation}
    \begin{split}
        \bar{\mathbf{x}}'_{ij} &= \bar{\mathbf{x}}_{ij}{\rm cos}(\theta) + \bar{\mathbf{y}}_{ij}{\rm sin}(\theta) \\
        \bar{\mathbf{y}}'_{ij} &= \bar{\mathbf{y}}_{ij}{\rm cos}(\theta) - \bar{\mathbf{x}}_{ij}{\rm sin}(\theta)
    \end{split}
    \label{eq:align_vecs}
\end{equation}
within their corresponding branching plane, to ensure maximal overlap between the rotated $\bar{\mathbf{x}}'_{ij}$ vector in Eq.~\eqref{eq:align_vecs} and the (unrotated) reference branching space vector of XMS-CASPT2, denoted here as $\bar{\mathbf{x}}_{ij}^{\rm ref}$. This requires $\theta$ in Eq.~\eqref{eq:align_vecs} to take the value
\begin{equation}
    \theta = {\rm arctan}\left(\frac{\bar{\mathbf{y}}_{ij} \cdot \bar{\mathbf{x}}_{ij}^{\rm ref}}{\bar{\mathbf{x}}_{ij} \cdot \bar{\mathbf{x}}_{ij}^{\rm ref}}\right) \,\,\, ,
    \label{eq:theta}
\end{equation}
where $\theta = \theta_0 + m\pi$ for $m = 0, \pm1, \pm 2, ...$ and $\theta_0$ is the initial value solution of Eq.~\eqref{eq:theta} as stated in Ref. \onlinecite{liu_analytical_2021}. (Note again, this rotation procedure does not change the branching plane obtained by the given electronic structure method of interest, but simply modifies the orientation of the branching space vectors within the plane by a rigid rotation.)\cite{ferre_description_2015} The value of $m$ was chosen to ensure that $\bar{\mathbf{x}}'_{ij} \cdot \bar{\mathbf{x}}_{ij}^{\rm ref} > 0$. Therefore, in all cases $m$ took a value of 0 or 1. Equally, the rotated $\bar{\mathbf{y}}'_{ij}$ vector obtained in Eq. \ref{eq:align_vecs} agrees well, in general, with the unrotated $\bar{\mathbf{y}}_{ij}^{\rm ref}$ vector of XMS-CASPT2 (up to a trivial negative sign).\cite{liu_analytical_2021}

As a result, a given geometry within the 29$\times$29 grid used to construct the various branching planes studied in this work can be defined as,\cite{fdez_galvan_analytical_2016}
\begin{equation}
    \R(\bar{x}_{ij}, \bar{y}_{ij}) = \R_{\rm CX} + \bar{x}_{ij}\bar{\mathbf{x}}_{ij} + \bar{y}_{ij}\bar{\mathbf{y}}_{ij}
\end{equation}
for XMS-CASPT2 and defined as,
\begin{equation}
    \R(\bar{x}'_{ij}, \bar{y}'_{ij}) = \R_{\rm CX} + \bar{x}'_{ij}\bar{\mathbf{x}}'_{ij} + \bar{y}'_{ij}\bar{\mathbf{y}}'_{ij}
\end{equation}
for all other electronic structure methods, where $\R_{\rm CX}$ is the optimised MECX (or MECP) geometry, and $\bar{x}_{ij}$/$\bar{y}_{ij}$ and $\bar{x}'_{ij}$/$\bar{y}'_{ij}$ are arbitrary lengths (in units of Å) along the corresponding orthonormalised (and linearised)\cite{carpenter_conical_2022} branching space vectors given in Eqs.~\eqref{eq:unalign_vecs} and \eqref{eq:align_vecs}, respectively. (Note, $\bar{x}_{ij}$/$\bar{y}_{ij}$ and $\bar{x}'_{ij}$/$\bar{y}'_{ij}$ should not be confused with the magnitudes of the orthonormalised branching space vectors themselves, which are trivially unity by definition, i.e., here $\bar{x}_{ij} \ne ||\bar{\mathbf{x}}_{ij}||$, $\bar{y}_{ij} \ne ||\bar{\mathbf{y}}_{ij}||$ etc.) It is these lengths ($\bar{x}_{ij}$/$\bar{y}_{ij}$ and $\bar{x}'_{ij}$/$\bar{y}'_{ij}$) that we label the x- and y-axes with in the branching space plots studied in this work.

\subsubsection{CX Branching space topography parameters}

To provide a numerical comparison of the topography of the MECXs obtained by different electronic structure methods in this work, we calculated the CX topography parameters, $\mathcal{P}$ and $\mathcal{B}$, as defined in Eqs. (57) and (58) of Ref. \onlinecite{fdez_galvan_analytical_2016}, respectively. For reference, the MECXs are characterised as peaked ($\mathcal{P} < 1$) or sloped ($\mathcal{P} > 1$), and bifurcating ($\mathcal{B} < 1$) or single-path ($\mathcal{B} > 1$) -- see Table \ref{tab:topog_par} for the $\mathcal{P}$ and $\mathcal{B}$ values and MECX characterisations. To be consistent with Ref. \onlinecite{fdez_galvan_analytical_2016}, we chose the value of $n$ in the definition of $\beta$ above, such that the value of the asymmetry parameter, $\Delta_{gh}$, (given in Eq. (45) of Ref. \onlinecite{fdez_galvan_analytical_2016}) was greater or equal to 0. Therefore, in all cases $n$ took a value of 0 or 1.

\begin{table}[h!]
    \centering
    \caption{CX branching space topography parameters as defined in Eqs. (57) and (58) in Ref. \onlinecite{fdez_galvan_analytical_2016}.}
        \begin{tabular}{lrrll}
        \hline
         & $\mathcal{P}$ & $\mathcal{B}$ & Characterisation & Fig.\\
        \hline
        \textbf{protonated formaldimine S$_\mathbf{2}$/S$_\mathbf{1}$ MECX} & & & \\
        XMS(3)-CASPT2(6/4) & 0.02 & 0.29 & peaked bifurcating & 2(a) \\
        ADC(2) & 0.08 & 0.45 & peaked bifurcating & 2(b) \\
        \vspace{0.1in}
        LR-TDDFT/TDA/PBE0 & 0.59 & 0.86 & peaked bifurcating & 2(c) \\
        \textbf{protonated formaldimine S$_\mathbf{1}$/S$_\mathbf{0}$ MECX} & & & \\
        \vspace{0.1in}
        XMS(3)-CASPT2(6/4) & 1.49 & 1.32 & sloped single-path & 3(a) \\
        \textbf{pyrazine S$_\mathbf{2}$/S$_\mathbf{1}$ MECX} & & & \\
        XMS(3)-CASPT2(10/8) & 3.57 & 1.96 & sloped single-path & 5(a) \\
        ADC(2) & 12.78 & 1.14 & sloped single-path & 5(b) \\
        LR-TDDFT/TDA/PBE0 & 7.16 & 2.64 & sloped single-path & 5(c) \\
        LR-TDDFT/TDA/LC-$\omega$PBE & 5.19 & 2.60 & sloped single-path & S7 \\
        \hline
        \end{tabular}
        \label{tab:topog_par}
    \label{tab:topography_parameters}
\end{table}


\subsubsection{Practicalities of plotting the extended S$_1$/S$_0$ branching space in (LR-TD)DFT/TDA}

We note that we experienced initial issues with KS SCF convergence for certain geometries along the extended S$_1$/S$_0$ branching plane. Such convergence problems have been attributed previously by Casida and coworkers\cite{huix-rotllant_assessment_2010, ferre_description_2015, marques_non-bornoppenheimer_2012, casida_progress_2012, tapavicza_mixed_2008} to an effective violation of non-interacting $v$-representability near the CX (i.e., the LUMO energy becomes lower than the HOMO energy). We followed Ref. \onlinecite{tapavicza_mixed_2008} and used the converged KS orbitals from neighbouring geometries as the initial guesses for the KS SCF calculations at the problematic geometries. This lead to much improved convergence and smoother PESs along the extended S$_1$/S$_0$ branching plane.

\clearpage
\section{Supplementary figures}

\begin{figure}[h!]
    \centering
    \includegraphics[width=0.75\textwidth]{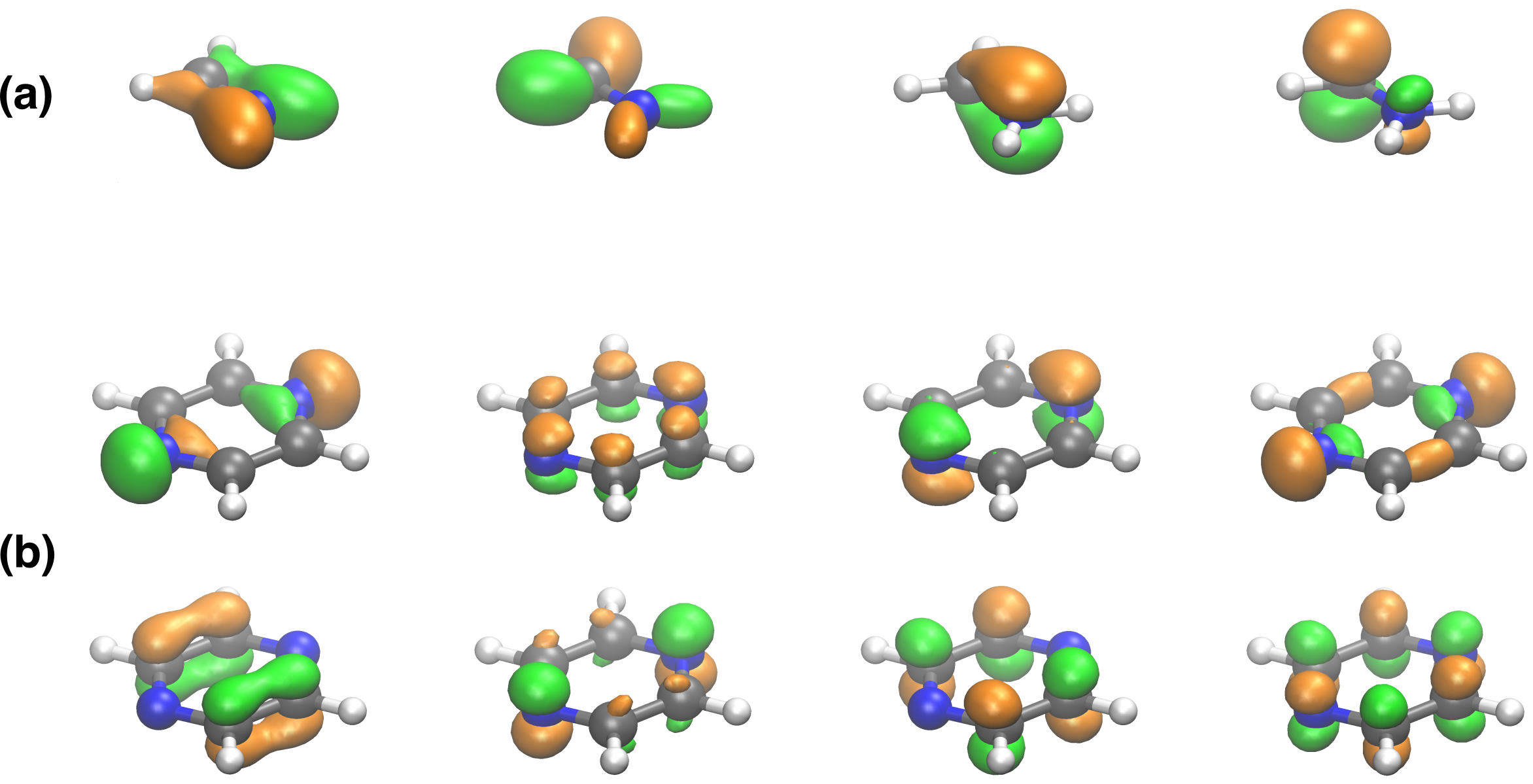}
    \caption{Underlying SA-CASSCF natural orbitals employed in (a) the XMS(3)-CASPT2(6/4)/cc-pVTZ calculation for protonated formaldimine and (b) the XMS(3)-CASPT2(10/8)/cc-pVTZ calculations for pyrazine. The active space orbitals are presented here for the S$_0$ minimum geometry of the two respective molecules plotted using an isovalue of 0.1.}
    \label{fig:ESI_active_space_mos}
\end{figure}

\begin{figure}[h!]
    \centering
    \includegraphics[width=0.45\textwidth]{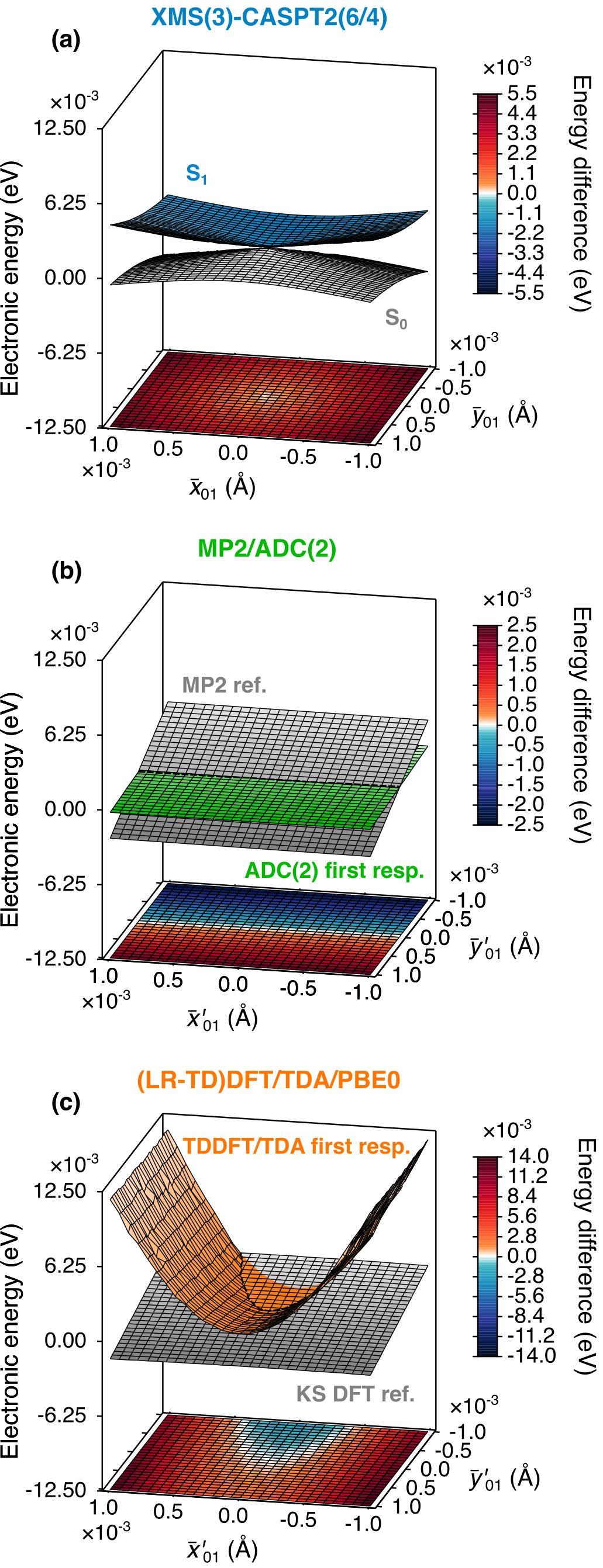}
    \caption{Branching space of the S$_1$/S$_0$ MECX (or MECP) in protonated formaldimine. Same as in Fig. 3 except all plots are plotted at the same electronic energy range on the z-axis.}
    \label{fig:ESI_pro_form_s1s0_bp_same_z-axis}
\end{figure}

\begin{figure}[h!]
    \centering
    \includegraphics[width=0.45\textwidth]{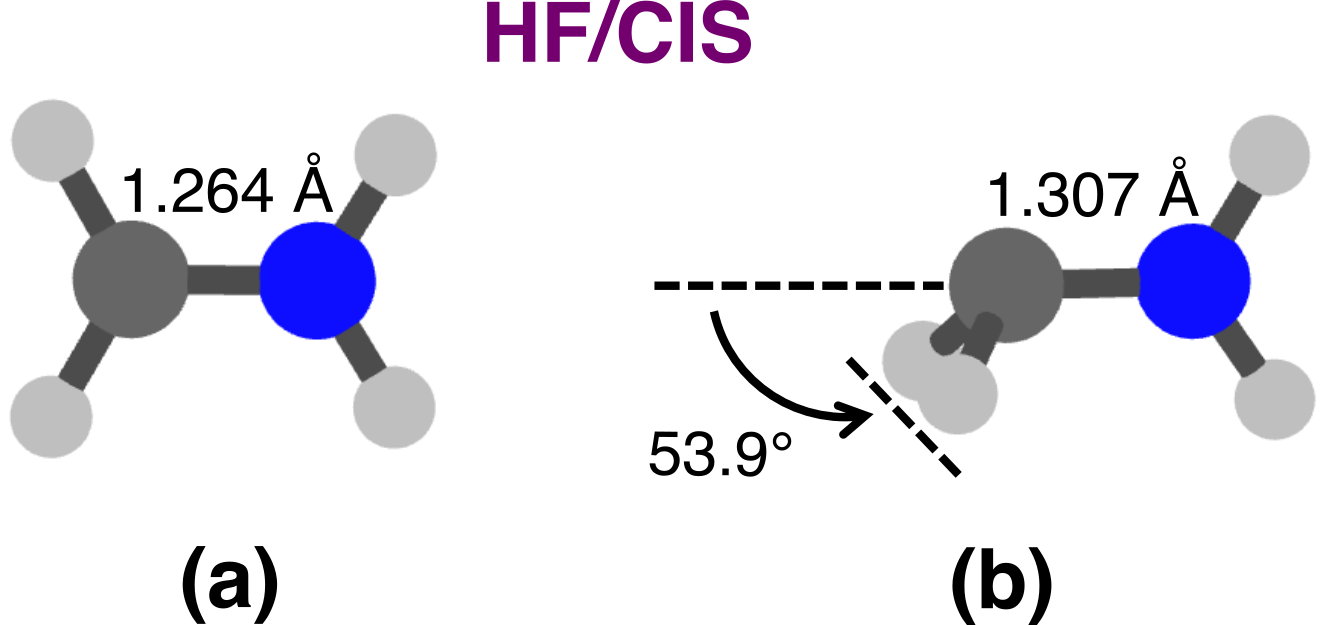}
    \caption{Molecular structure of (a) the S$_0$ minimum and (b) the S$_2$/S$_1$ MECX geometries in protonated formaldimine optimised at the HF/CIS/cc-pVDZ levels of theory.}
    \label{fig:ESI_pro_form_hf-cis_geoms}
\end{figure}

\begin{figure}[h!]
    \centering
    \includegraphics[width=0.55\textwidth]{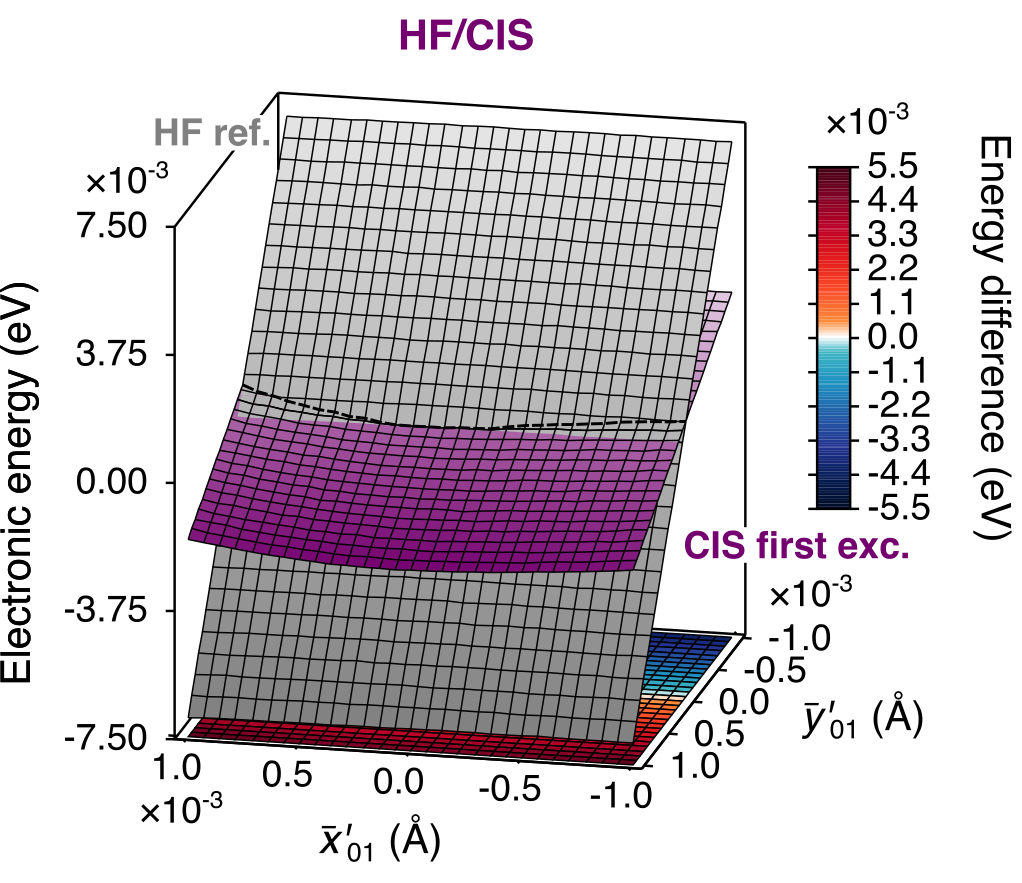}
    \caption{Branching space of the S$_1$/S$_0$ MECP in protonated formaldimine. Reference and first excited states obtained with HF/CIS/cc-pVDZ (purple/grey). The MECP geometry and branching space vectors were obtained at the same level of theory used to compute the PESs. The dashed line in the plot indicates the seam where $E_{1}(\R) = E_{0}(\R)$. (We note that the rendering of the colours for the PESs does not reflect precisely this intersection.) The base shows a 2D colour map of the energy difference between the reference and first excited states (see colour bar on the right).}
    \label{fig:ESI_pro_form_CIS_S1S0_branch_space}
\end{figure}

\begin{figure}[h!]
    \centering
    \includegraphics[width=0.65\textwidth]{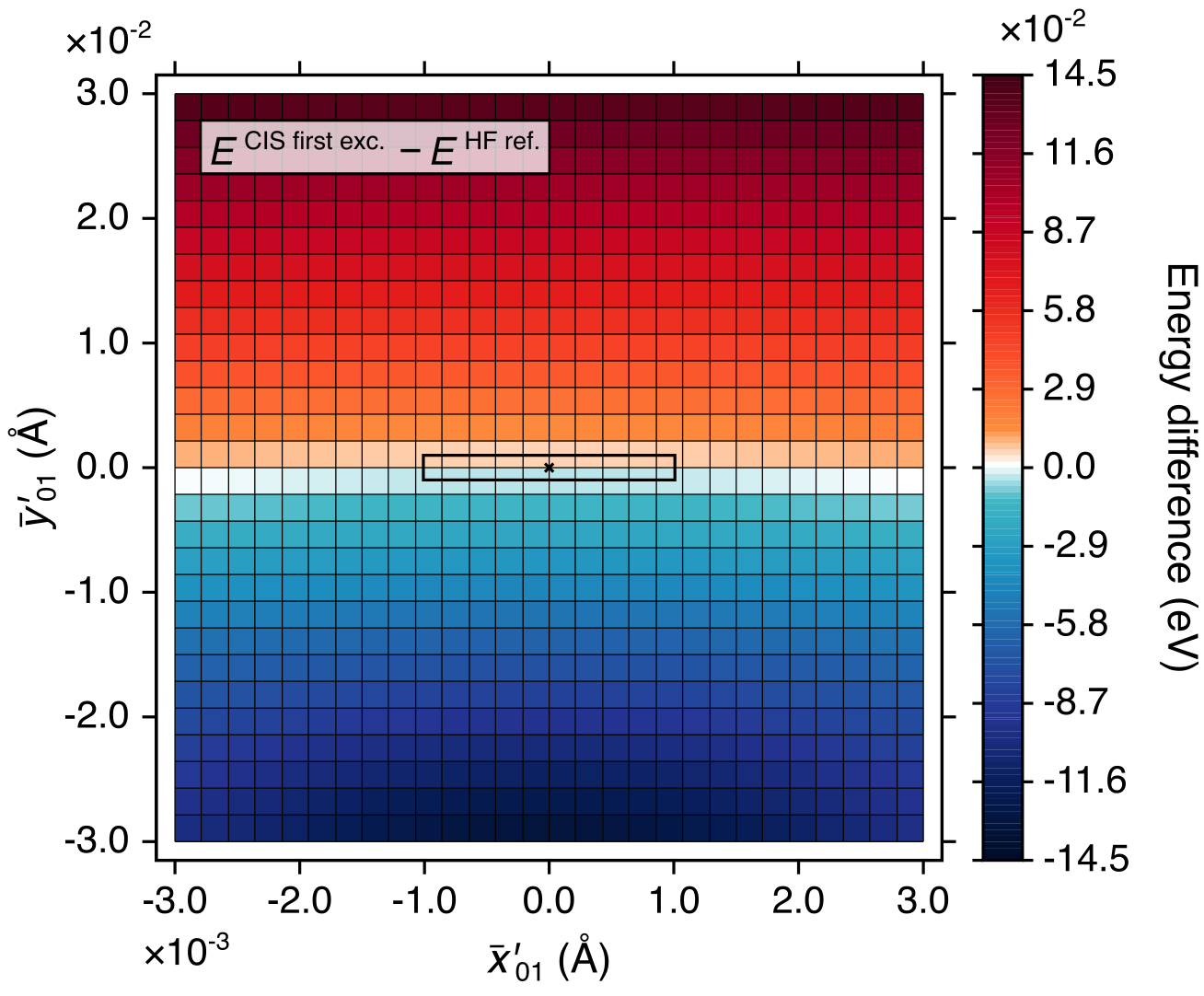}
    \caption{2D map of the energy difference between the reference and first excited states obtained with HF/CIS/cc-pVDZ in the vicinity of the S$_1$/S$_0$ MECP along an extended branching plane ($\pm0.003 \times \bar{\mathbf{x}}'_{ij}(\R)$, $\pm0.03 \times \bar{\mathbf{y}}'_{ij}(\R)$). The black box encloses the area spanned by the branching plane used to generate the plots in Fig. \ref{fig:ESI_pro_form_CIS_S1S0_branch_space} ($\pm0.001 \times \bar{\mathbf{x}}'_{ij}(\R)$, $\pm0.001 \times \bar{\mathbf{y}}'_{ij}(\R)$); the black cross indicates the location of the MECP geometry.}
    \label{fig:ESI_pro_form_CIS_S1S0_zoom_out}
\end{figure}

\begin{figure}[h!]
    \centering
    \includegraphics[width=0.35\textwidth]{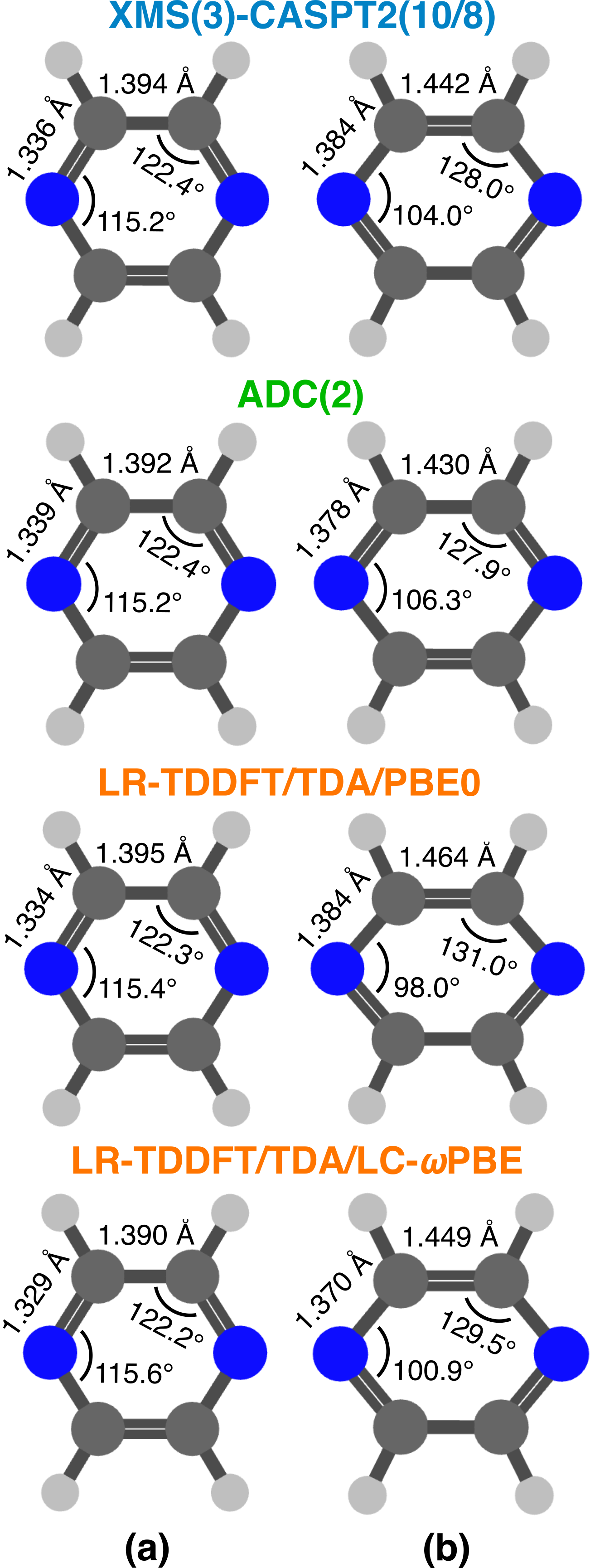}
    \caption{Molecular structure of (a) the S$_0$ minimum and (b) the S$_2$/S$_1$ MECX geometries in pyrazine optimised at the XMS(3)-CASPT2(10/8)/cc-pVTZ, ADC(2)/cc-pVTZ, LR-TDDFT/TDA/PBE0/cc-pVDZ and LR-TDDFT/TDA/LC-$\omega$PBE/cc-pVDZ levels of theory.}
    \label{fig:ESI_pyra_geoms}
\end{figure}

\begin{figure}[h!]
    \centering
    \includegraphics[width=0.55\textwidth]{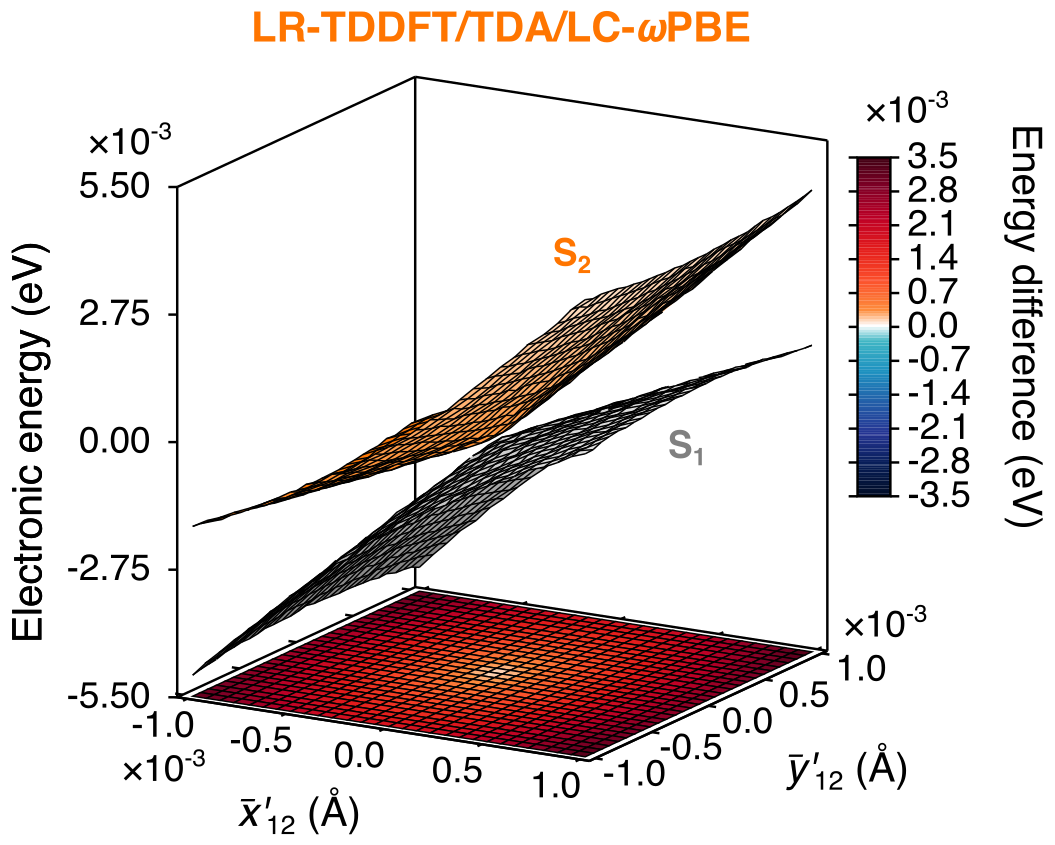}
    \caption{Branching space of the S$_2$/S$_1$ MECX in protonated formaldimine. S$_1$ and S$_2$ excited states obtained with LR-TDDFT/TDA/LC-$\omega$PBE/cc-pVDZ (orange/grey). The MECX geometry and branching space vectors were obtained at the same level of theory used to compute the PESs. The base shows a 2D colour map of the energy difference between the first and second excited states (see colour bar on the right).}
    \label{fig:ESI_pro_form_lc-wpbe_S1S0_branch_space}
\end{figure}

\clearpage
\bibliography{ref2}
\bibliographystyle{rsc}